\input harvmac
\let\includefigures=\iftrue
\let\useblackboard=\iftrue
\newfam\black 

\includefigures
\message{If you do not have epsf.tex (to include figures),}
\message{change the option at the top of the tex file.}
\input epsf
\def\figin{\epsfcheck\figin}\def\figins{\epsfcheck\figins}
\def\epsfcheck{\ifx\epsfbox\UnDeFiNeD
\message{(NO epsf.tex, FIGURES WILL BE IGNORED)}
\gdef\figin##1{\vskip2in}\gdef\figins##1{\hskip.5in}
\else\message{(FIGURES WILL BE INCLUDED)}%
\gdef\figin##1{##1}\gdef\figinbs##1{##1}\fi}
\def\DefWarn#1{}
\def\figinsert{\goodbreak\midinsert}
\def\ifig#1#2#3{\DefWarn#1\xdef#1{fig.~\the\figno}
\writedef{#1\leftbracket fig.\noexpand~\the\figno}%
\figinsert\figin{\centerline{#3}}\medskip\centerline{\vbox{
\baselineskip12pt\advance\hsize by -1truein
\noindent\footnotefont{\bf Fig.~\the\figno:} #2}}
\endinsert\global\advance\figno by1}
\else
\def\ifig#1#2#3{\xdef#1{fig.~\the\figno}
\writedef{#1\leftbracket fig.\noexpand~\the\figno}%
\global\advance\figno by1} \fi

\def\id{{1 \kern-.28em {\rm l}}}

\def\K3{{\bf K3}}
\def\journal#1&#2(#3){\unskip, \sl #1\ \bf #2 \rm(19#3) }
\def\andjournal#1&#2(#3){\sl #1~\bf #2 \rm (19#3) }

\def\bar{\overline}
\def\hat{\widehat}
\def\ie{{\it i.e.}}
\def\eg{{\it e.g.}}

\def\tilde{\widetilde}

\def\frac#1#2{{#1\over#2}}

\def\half{\frac12}

\def\inbar{\,\vrule height1.5ex width.4pt depth0pt}
\def\IC{\relax\hbox{$\inbar\kern-.3em{\rm C}$}}
\def\IR{\relax{\rm I\kern-.18em R}}
\def\IZ{\relax{\rm I\kern-.18em Z}}

%
%

%
\catcode`\@=11
\def\slash#1{\mathord{\mathpalette\c@ncel{#1}}}
\overfullrule=0pt
\def\AA{{\cal A}}

\def\DD{{\cal D}}

\def\FF{{\cal F}}

\def\LL{{\cal L}}

\def\WW{{\cal W}}

\def\underrel#1\over#2{\mathrel{\mathop{\kern\z@#1}\limits_{#2}}}

\catcode`\@=12


%

\def\exp{{\rm exp}}


\def\ie{{\it i.e.}}
\def\eg{{\it e.g.}}

\lref\WessCP{
  J.~Wess and J.~Bagger,
  ``Supersymmetry and supergravity,''
Princeton, USA: Univ. Pr. (1992) 259 p.
}

\lref\AlmuhairiWS{
  A.~Almuhairi and J.~Polchinski,
  ``Magnetic $AdS x R^2$: Supersymmetry and stability,''
[arXiv:1108.1213 [hep-th]].
}

\lref\WittenYC{
  E.~Witten,
  ``Phases of N=2 theories in two-dimensions,''
Nucl.\ Phys.\ B {\bf 403}, 159 (1993).
[hep-th/9301042].
}

\lref\CremadesWA{
  D.~Cremades, L.~E.~Ibanez and F.~Marchesano,
  ``Computing Yukawa couplings from magnetized extra dimensions,''
JHEP {\bf 0405}, 079 (2004).
[hep-th/0404229].
}

\lref\GiveonSR{
  A.~Giveon and D.~Kutasov,
  ``Brane dynamics and gauge theory,''
Rev.\ Mod.\ Phys.\  {\bf 71}, 983 (1999).
[hep-th/9802067].
}

\lref\BrodieSZ{
  J.~H.~Brodie and A.~Hanany,
  ``Type IIA superstrings, chiral symmetry, and N=1 4-D gauge theory dualities,''
Nucl.\ Phys.\ B {\bf 506}, 157 (1997).
[hep-th/9704043].
}

\lref\BerkoozKM{
  M.~Berkooz, M.~R.~Douglas and R.~G.~Leigh,
  ``Branes intersecting at angles,''
Nucl.\ Phys.\ B {\bf 480}, 265 (1996).
[hep-th/9606139].
}

\lref\BarbonZU{
  J.~L.~F.~Barbon,
  ``Rotated branes and N=1 duality,''
Phys.\ Lett.\ B {\bf 402}, 59 (1997).
[hep-th/9703051].
}

\lref\HananyIE{
  A.~Hanany and E.~Witten,
  ``Type IIB superstrings, BPS monopoles, and three-dimensional gauge dynamics,''
Nucl.\ Phys.\ B {\bf 492}, 152 (1997).
[hep-th/9611230].
}

\lref\ElitzurFH{
  S.~Elitzur, A.~Giveon and D.~Kutasov,
  ``Branes and N=1 duality in string theory,''
Phys.\ Lett.\ B {\bf 400}, 269 (1997).
[hep-th/9702014].
}

\lref\TerningBQ{
  J.~Terning,
  ``Modern supersymmetry: Dynamics and duality,''
(International series of monographs on physics. 132).
}

\lref\TongQJ{
  D.~Tong,
  ``The Quantum Dynamics of Heterotic Vortex Strings,''
JHEP {\bf 0709}, 022 (2007).
[hep-th/0703235 [HEP-TH]].
}

\lref\AharonyGP{
  O.~Aharony,
  ``IR duality in d = 3 N=2 supersymmetric USp(2N(c)) and U(N(c)) gauge theories,''
Phys.\ Lett.\ B {\bf 404}, 71 (1997).
[hep-th/9703215].
}

\lref\GiveonZN{
  A.~Giveon and D.~Kutasov,
  ``Seiberg Duality in Chern-Simons Theory,''
Nucl.\ Phys.\ B {\bf 812}, 1 (2009).
[arXiv:0808.0360 [hep-th]].
}

\lref\BeniniCZ{
  F.~Benini and N.~Bobev,
  ``Exact two-dimensional superconformal R-symmetry and c-extremization,''
Phys.\ Rev.\ Lett.\  {\bf 110}, no. 6, 061601 (2013).
[arXiv:1211.4030 [hep-th]].
}
\lref\BeniniCDA{
  F.~Benini and N.~Bobev,
  ``Two-dimensional SCFTs from wrapped branes and c-extremization,''
JHEP {\bf 1306}, 005 (2013).
[arXiv:1302.4451 [hep-th]].
}

\lref\SeibergPQ{
  N.~Seiberg,
  ``Electric - magnetic duality in supersymmetric nonAbelian gauge theories,''
Nucl.\ Phys.\ B {\bf 435}, 129 (1995).
[hep-th/9411149].
}

\lref\WittenYU{
  E.~Witten,
  ``On the conformal field theory of the Higgs branch,''
JHEP {\bf 9707}, 003 (1997).
[hep-th/9707093].
}

\lref\IntriligatorEX{
  K.~A.~Intriligator and N.~Seiberg,
  ``Mirror symmetry in three-dimensional gauge theories,''
Phys.\ Lett.\ B {\bf 387}, 513 (1996).
[hep-th/9607207].
}

\lref\IntriligatorAU{
  K.~A.~Intriligator and N.~Seiberg,
  ``Lectures on supersymmetric gauge theories and electric - magnetic duality,''
Nucl.\ Phys.\ Proc.\ Suppl.\  {\bf 45BC}, 1 (1996).
[hep-th/9509066].
}

\lref\AharonyDHA{
  O.~Aharony, S.~S.~Razamat, N.~Seiberg and B.~Willett,
  ``3d dualities from 4d dualities,''
JHEP {\bf 1307}, 149 (2013).
[arXiv:1305.3924 [hep-th]].
}

\lref\ShifmanKJ{
  M.~Shifman and A.~Yung,
  ``Large-N Solution of the Heterotic N=(0,2) Two-Dimensional CP(N-1) Model,''
Phys.\ Rev.\ D {\bf 77}, 125017 (2008), [Erratum-ibid.\ D {\bf 81}, 089906 (2010)].
[arXiv:0803.0698 [hep-th]].
}

\lref\GaddeLXA{
  A.~Gadde, S.~Gukov and P.~Putrov,
  ``(0,2) Trialities,''
[arXiv:1310.0818 [hep-th]].
}

\lref\KarndumriIQA{
  P.~Karndumri and E.~O Colgain,
Phys.\ Rev.\ D {\bf 87}, 101902 (2013).
[arXiv:1302.6532 [hep-th]].
}

\lref\KutasovHHA{
  D.~Kutasov and J.~Lin,
[arXiv:1401.5558 [hep-th]].
}

\Title{}
{\vbox{\centerline{(0,2) Dynamics From Four Dimensions}
\bigskip
}}
\bigskip

\centerline{\it  David Kutasov and Jennifer Lin}
\bigskip
\centerline{EFI and Department of Physics, University of
Chicago} \centerline{5640 S. Ellis Av., Chicago, IL 60637, USA }
\smallskip

\vglue .3cm

\bigskip

\let\includefigures=\iftrue
\bigskip
\noindent
We study $(0,2)$ supersymmetric two-dimensional theories obtained by compactifying four-dimensional $N=1$ supersymmetric theories on a two-torus, with a magnetic field for a global $U(1)$ symmetry, and present evidence that Seiberg duality in four dimensions leads to an identification of different models of this type.
\bigskip

\Date{}

\newsec{Introduction}

Supersymmetric quantum field theories (SQFTs) with four supercharges in various dimensions $(d\le 4)$ have been extensively studied in the last few decades. Many of their properties have been elucidated using symmetries, anomalies, holomorphy, strong-weak coupling duality and related ideas (see \eg\ \refs{\IntriligatorAU,\TerningBQ} for reviews of theories in $d=4$), as well as insights from string theory, where many such theories can be realized as low energy theories on branes \refs{\HananyIE,\ElitzurFH,\GiveonSR}.   An interesting aspect of this program was the discovery of connections between the dynamics of such theories in different dimensions (see \eg\ \AharonyDHA\ for a recent discussion of the $3d-4d$ connection). 

It is natural to ask how much of this progress can be extended to theories with two supercharges in $d\le 3$. In three dimensions, such theories have  the minimal amount of supersymmetry, $N=1$, for which many of the techniques mentioned above are inapplicable. In two dimensions, one can consider either $(1,1)$ SQFTs, which suffer from the same problem,\foot{Indeed, one way to construct $(1,1)$ SQFTs in two dimensions is to consider the low energy limit of three dimensional $N=1$ SQFTs on a circle.} or models with chiral, say $(0,2)$, SUSY which seem more promising from this point of view, due to their inherent chirality and the fact that they have the minimal amount of supersymmetry for which holomorphy, anomalies etc. place strong constraints on the dynamics. They are also interesting for applications to (heterotic worldsheet) string theory.  

In order to explore the $2d-4d$ connection for theories with $(0,2)$ supersymmetry, we need a way to associate a $(0,2)$ model to a given four dimensional $N=1$ SQFT. One way to do this in a large class of examples is the following. Given a four dimensional $N=1$ SQFT with a global (non R) symmetry, one can couple the current supermultiplet to an external vector superfield, which consists of a vector field $A_\mu$, a gaugino $\lambda_\alpha$, and an auxiliary field $D$. These fields are non-dynamical, but can still take non-zero expectation values. In a spacetime of the form $\IR^{1,1}\times {\rm T}^2$, labeled by the coordinates $(x^0, x^3)\in \IR^{1,1}$, $(x^1,x^2)\in {\rm T}^2$, it is natural to turn on an expectation value for the magnetic field through the torus, 
\eqn\magfield{F_{12}=B~.}
This breaks supersymmetry completely, as can be seen by examining the  variation of the external gaugino field: $\delta\lambda = F_{\mu\nu}\sigma^{\mu\nu}\epsilon$, with $\mu,\nu=0,1,2,3$. For a finite magnetic field, $\delta\lambda$ is non-zero for all $\epsilon$. To preserve some of the supersymmetry we also turn on a non-zero $D$ field \AlmuhairiWS, which modifies the gaugino variation to
\eqn\gaugino{
\delta\lambda = (F_{\mu\nu}\sigma^{\mu\nu} + iD)\epsilon~.
}
Plugging \magfield\ into \gaugino, one finds
\eqn\bandd{\delta\lambda =i\left(\matrix{D-B & 0 \cr 0 & D+B} \right)\left(\matrix{\epsilon_- \cr \epsilon_+} \right)~,}
where $\epsilon_-$ $(\epsilon_+)$ generates right (left) moving supersymmetry on $\IR^{1,1}$. For generic $B$ and $D$, supersymmetry is broken as before, but for $D = \pm B$, some of it remains unbroken. For $D=B$, the r.h.s. of \bandd\ vanishes for all $\epsilon_-$ (and $\epsilon_+=0$); $D=-B$ is the same with $\epsilon_+\leftrightarrow \epsilon_-$. In other words, for $D=B$ the theory preserves $(0,2)$ SUSY while $D = -B$ gives a theory with $(2,0)$ SUSY. Without loss of generality we can focus on the $(0,2)$ case. 

The above construction can be alternatively interpreted in terms of the three dimensional $N=2$ SQFT obtained by compactifying the original four dimensional $N=1$ supersymmetric theory on a circle. The three dimensional theory has the $U(1)$ global symmetry of the underlying four dimensional model. To get a $(0,2)$ SQFT in $1+1$ dimensions, we compactify this theory on one more circle, and turn on a real mass term associated with the $U(1)$ symmetry, that depends on position along the circle. This point of view will be useful below. 

The procedure that associates a two dimensional $(0,2)$ model to a four dimensional $N=1$ SQFT with a global $U(1)$ symmetry is clearly not unique in theories with large global symmetry groups, since then there are many inequivalent ways to choose the $U(1)$ current that figures in the construction. Also, it may be that the resulting $(0,2)$ theory breaks supersymmetry when quantum effects are taken into account.  The purpose of this paper is to study this class of theories, and address these and other issues. We will see that for some choices of the $U(1)$ symmetry,  supersymmetry is broken in the quantum theory, while for others it is not. Even in cases where SUSY is unbroken, quantum effects play an important role in the dynamics.

We will also discuss the dependence of the properties of the two dimensional $(0,2)$ theory on those of the underlying four dimensional one. For example, if the higher-dimensional theory exhibits Seiberg duality \SeibergPQ, one can ask whether the two dimensional theories obtained by compactifying the electric and magnetic models inherit it. On the one hand, one can argue that they should, since no phase transitions are expected as a function of the parameters that the models depend on. On the other hand, one might worry that the duality might be spoiled by the presence of scalar fields in the adjoint representation (coming from dimensional reduction of the four dimensional gauge field), which are sensitive to the rank of the gauge group.  We shall argue that the duality survives in two dimensions, and explain how it deals with the adjoint fields. Similarly, one can ask how mirror symmetry in three dimensions \IntriligatorEX\ is realized after the reduction. We leave a discussion of this issue to another publication.

Although the construction described above is general, we shall mostly focus on the case of $N=1$ supersymmetric QCD with gauge group $U(N_c)$ and matter in the (anti) fundamental representation. This class of theories can be realized in string theory as low energy theories of systems of $D$-branes and $NS5$-branes \refs{\HananyIE,\ElitzurFH,\GiveonSR}, and we shall find this description useful for our purposes. 

The plan of the paper is as follows. In section 2 we briefly review the basic structure of $(0,2)$ supersymmetric gauge theories in two dimensions. In section 3 we describe the effects of background magnetic and $D$ fields on the spectrum of free charged four dimensional superfields compactified on a torus. In section 4 we include interactions and discuss the effect of background fields on a four dimensional $N=1$ supersymmetric gauge theory, focusing on $N=1$ SQCD with gauge group $U(N_c)$ and fundamental matter. In section 5 we embed this theory in string theory as a low energy theory on branes, and explain the effects of the background fields on it. This embedding is known to be useful for studying the classical and quantum low energy dynamics of various gauge theories in different dimensions \GiveonSR, and this turns out to be the case here as well. 

In section 6 we describe our construction from the point of view of the three dimensional $N=2$ supersymmetric gauge theory obtained by compactifying four dimensional $N=1$ SQCD on a circle. In the three dimensional description, the magnetic field for the background $U(1)$ becomes a real mass term for the fundamentals that depends linearly on one of the spatial directions. This description is useful for analyzing the dynamics, particularly in the brane realization, since the background fields correspond to geometric deformations in the extra dimensions. 

In section 7 we discuss some properties of the two dimensional quantum theories obtained from our construction. We show that the classical Coulomb moduli space is lifted in the quantum theory, and is replaced by a discrete set of vacua. We describe the low energy theory in each of these vacua and argue that four dimensional theories related by Seiberg duality give rise to the same low energy $(0,2)$ theories in two dimensions.  Section 8 contains a brief discussion of our results. Some technical details appear in an appendix.

\newsec{(0,2) supersymmetry in two dimensions}

In this section we review some aspects of $(0,2)$ supersymmetric field theories in $1+1$ dimensions which will play a role in our discussion below. Our main goal is to establish the notation, which follows that of \WittenYC. 

\subsec{General structure}

In addition to the bosonic coordinates $(x^0,x^3)$, $(0,2)$ superspace has two fermionic coordinates $(\theta^+, \bar\theta^+)$.  The  right-moving supercharges act on superspace as follows:
\eqn\q{
Q_+ = \frac\partial{\partial\theta^+} + i\bar\theta^+ (\partial_0 + \partial_3)~, \qquad \bar Q_+ = -\frac{\partial}{\partial\bar\theta^+} - i\theta^+(\partial_0 + \partial_3)~.
}
They satisfy $Q_+^2=\bar Q_+^2=0$, $\{Q_+,\bar Q_+\}=-2i(\partial_0 + \partial_3)$. The supercharges anticommute with the superspace covariant derivatives
\eqn\Dderiv{
D_+ = \frac{\partial}{\partial\theta^+} - i\bar\theta^+(\partial_0 + \partial_3)~, \qquad
\bar D_+ = -\frac\partial{\partial\bar\theta^+} + i\theta^+(\partial_0 + \partial_3)~,
}
which satisfy $D_+^2 = \bar D_+^2 = 0$, $\{D_+,\bar D_+\}=2i(\partial_0 + \partial_3)$.

To construct $(0,2)$ supersymmetric gauge theory, we extend the superspace derivatives $D_+, \bar D_+$, $\partial_0$, $\partial_3$ to gauge covariant superderivatives $\DD_+, \bar \DD_+, \DD_0, \DD_3$. These can be writtten in a Wess-Zumino type gauge as
\eqn\Dcov{\eqalign{
&\DD_0 + \DD_3 = \partial_0 + \partial_3 + i(A_0 + A_3)~, \cr
&\DD_+ = \frac\partial{\partial\theta^+} - i\bar\theta^+(\DD_0 + \DD_3)~, \cr
&\bar\DD_+ = -\frac\partial{\partial\bar\theta^+} + i\theta^+(\DD_0 + \DD_3)~, \cr
&\DD_0 - \DD_3 = \partial_0 - \partial_3 + iV~,
}}
where  
\eqn\vdef{
V = A_0 - A_3 - 2i\theta^+\bar\lambda_- - 2i\bar\theta^+\lambda_- + 2\theta^+\bar\theta^+ D
}
is a vector superfield which transforms in the adjoint representation of the gauge group $G$. $\lambda_-$ is the left-moving gaugino, while $D$ is a non-propagating auxiliary field. These fields and the gauge field can be combined into the $(0,2)$  field strength
\eqn\upsilone{
\Upsilon = [\bar\DD_+, \DD_0 - \DD_3] = -2(\lambda_- - i\theta^+(D - iF_{03}) - i\theta^+\bar\theta^+(D_0 + D_3)\lambda_-)~,
}
which transforms in the adjoint representation of the gauge group, and satisfies the chirality constraint $\bar\DD_+\Upsilon = 0$.

We will focus on two types of matter superfields in a representation $r$ of the gauge group $G$. One is the bosonic chiral superfield $\Phi$ with component expansion
\eqn\phie{
\Phi = \phi + \sqrt{2}\theta^+\psi_+ - i\theta^+\bar\theta^+(D_0 + D_3)\phi~,
}
obeying $\bar \DD_+\Phi = 0$.  Here $\phi$ is a complex scalar field, and $\psi_+$ a complex right-moving fermion. The second is the Fermi superfield 
\eqn\lambdae{
\Lambda =  \psi_- - \sqrt{2}\theta^+\FF - i\theta^+\bar\theta^+(D_0 + D_3)\psi_- - \sqrt{2}\bar\theta^+ E~,
}
whose only propagating degree of freedom is a complex left-moving fermion $\psi_-$. $\Lambda$ obeys the superspace constraint
\eqn\dlambda{
\bar\DD_+\Lambda = \sqrt{2}E~, \qquad \bar\DD_+ E = 0~.
}
In \upsilone\ -- \lambdae, $D_0, D_3$ are the usual gauge-covariant derivatives $D_\mu = \partial_\mu + i A^a_\mu T^a$, or equivalently the superderivatives \Dcov\ evaluated at $\theta^+ = \bar\theta^+ = 0$. $T^a$ are the generators of $G$ in the representation $r$, and $E$ is a chiral superfield, usually taken to be a function of the basic chiral superfields in the theory \WittenYC.

The natural supersymmetric actions for the above superfields are
\eqn\saction{\eqalign{
S_\Upsilon &= \frac{1}{8g^2}\Tr \int d^2x d^2\theta  \bar\Upsilon\Upsilon~,  \cr
S_\Phi &= -\frac i2\int d^2xd^2\theta \bar\Phi(\DD_0 - \DD_3)\Phi~, \cr
S_\Lambda &= -\frac 12 \int d^2x d^2\theta \bar\Lambda\Lambda~,
}}
with component expansions
\eqn\scomponent{\eqalign{
S_\Upsilon &= \frac 1{g^2} \Tr \int d^2x\left\{\frac 12 F_{03}^2 + i\bar\lambda_-(D_0 + D_3)\lambda_- + \frac 12 D^2 \right\}~, \cr
S_\Phi &= \int d^2x\left\{-|D_\mu\phi|^2 + i\bar\psi_+(D_0 - D_3)\psi_+ - i\sqrt{2}\bar\phi T^a\lambda_-^a\psi_+ + i \sqrt{2}\phi T^a \bar\psi_+\bar\lambda^a_- + \bar\phi T^a\phi D^a  \right\}~, \cr
S_\Lambda &= \int d^2x \left\{i\bar\psi_-(D_0 + D_3)\psi_- + |\FF|^2 - |E|^2 - \left(\bar\psi_-\frac{\partial E}{\partial\phi_i}\psi_{+i} + \bar\psi_{+i}\frac{\partial\bar E}{\partial\bar\phi_i}\psi_-  \right)\right\}~.
}}
To include a non-trivial kinetic term for the $\Phi_i$, one can replace the  field $\bar\Phi_i$ in the action $S_{\Phi_i}$ by a more general Kahler form $K_i(\Phi,\bar\Phi)$. 

Other terms in the action that are often of interest have the general form $\int d^2x d\theta^+ (\dots)\left|_{\bar\theta^+ = 0} \right.$ + c.c. where $(\dots)$ is an anticommuting superfield annihilated by $\bar\DD_+$. One example is the Fayet-Iliopoulos (FI) term for a $U(1)$ symmetry,
\eqn\fiterm{
S_{\rm FI} = \frac t 4 \int d^2x d\theta^+\Upsilon\left|_{\bar\theta^+= 0}\right. + {\rm c.c.} = \frac{it}{2}\int d^2x(D - iF_{01}) + \rm{c.c.}~,
}
where
\eqn\fiparam{
t = ir + \frac{\theta}{2\pi}
}
is the complexified FI parameter. Another is the $(0,2)$ superpotential
\eqn\superpotential{
S_\WW = -\frac 1{\sqrt{2}}\int d^2x d\theta^+ \Lambda_a J^a\left|_{\bar\theta^+ = 0}\right.  + {\rm c.c.}
= -\int d^2x\left\{\FF_a J^a(\phi_i) + \psi_{-a}\psi_{+i}\frac{\partial J^a}{\partial\phi_i} \right\} + {\rm c.c.}~,
}
where $\Lambda_a$ are Fermi superfields, and $J^a$ are holomorphic functions of the (bosonic) chiral superfields.
Because of \dlambda, chirality of the superpotential \superpotential\ requires that
\eqn\wconstraint{
E \cdot J = 0~.
}

\subsec{Reduction of (2,2) superfields under (0,2) SUSY}

Before turning on the background $B$ and $D$ fields, the theories we shall study have $(2,2)$ supersymmetry in two dimensions. The background fields split the $(2,2)$ multiplets into $(0,2)$ ones.  Thus, it is useful to recall how $(2,2)$ superfields  decompose under $(0,2)$ supersymmetry.

$(2,2)$ SQFT is conveniently described in a superspace with coordinates $(x^\mu, \theta^\pm, \bar\theta^\pm)$, an obvious extension of the $(0,2)$ superspace described above to include the left-moving super-coordinates.  The right-moving supercovariant derivatives \Dcov\ are supplementd by  left-moving ones,
\eqn\Dderivm{
\DD_- = \frac{\partial}{\partial\theta^-} - i\bar\theta^-(\DD_0 - \DD_3)~,  \qquad \bar\DD_- = -\frac{\partial}{\partial\bar\theta^-} + i\theta^-(\DD_0 - \DD_3)~.
}
All the $(0,2)$ superfields described above fit naturally into two types of $(2,2)$ superfields. One is the chiral superfield $\Phi^{(2,2)}$, which satisfies the chirality constraints $\bar\DD_+\Phi^{(2,2)} = \bar\DD_-\Phi^{(2,2)} = 0$ and has the free field action
\eqn\ttphi{
S_\Phi^{(2,2)} = -\frac 14 \int d^2x d^4\theta \bar\Phi_i^{(2,2)}\Phi_i^{(2,2)}~.
}
The other is the twisted chiral superfield $\Sigma^{(2,2)}= \frac{1}{2\sqrt{2}}\{\bar\DD_+, \DD_-\}$, which describes the gauge field strength and obeys $\bar\DD_+\Sigma^{(2,2)} = \DD_-\Sigma^{(2,2)} = 0$. It has the action
\eqn\ttsigma{
S_g^{(2,2)} = -\frac{1}{4g^2}\int d^2xd^4\theta \bar\Sigma^{(2,2)}\Sigma^{(2,2)}~.
}
A $(2,2)$ chiral superfield splits into a $(0,2)$ chiral superfield and a  Fermi superfield, as can be seen from the $\theta^-$ expansion
\eqn\ttphid{
\Phi^{(2,2)} = \Phi^{(0,2)} + \sqrt{2}\theta^-\Lambda^{(0,2)} - i\theta^-\bar\theta^-(\DD_0 - \DD_3)\Phi^{(0,2)}~.
}
The $(2,2)$ field strength splits into an adjoint chiral superfield and a $(0,2)$ field strength superfield, 
\eqn\ttsigmad{
\Sigma^{(2,2)} = \Sigma^{(0,2)} +\frac i{\sqrt{2}}\bar\theta^-\Upsilon^{(0,2)} - i\theta^-\bar\theta^-(\DD_0 - \DD_3)\Sigma^{(0,2)}.
}
To reproduce the action of a $(2,2)$ chiral superfield that transforms non-trivially under a gauge symmetry, one must include a non-zero $E$ \lambdae\ for the Fermi superfield in \ttphid,
\eqn\ttphie{E = i\sqrt{2}\Sigma^aT^a\Phi~.}

\newsec{Free fields in a magnetic field}

In this section we review the spectrum of charged four dimensional free fields of spin $0$ and $1/2$ in the presence of a constant external magnetic field. We start with the non-compact case, and then discuss compactification on a two-torus. 


Consider a free massless (complex) scalar $\phi$ of charge $e$ under a $U(1)$ gauge field $A_\mu$. To study its dynamics in the background magnetic field \magfield, we turn on a background gauge field $A_2 = Bx_1$. The Klein-Gordon equation for $\phi$ then takes the form 
\eqn\scalareom{
(-\partial_0^2 + \partial_3^2 + \partial_1^2 + \tilde D_2^2)\phi = 0~,
}
where $\tilde D_2 = \partial_2 + ieBx_1$. The $1+1$ dimensional spectrum is obtained by writing
\eqn\phiexpand{
\phi(x^0, x^1, x^2, x^3) =\varphi(x^0, x^3)\chi(x^1, x^2)~.
}
If we take $\chi$ to be an eigenfunction of 
\eqn\hamiltonianB{
H = -(\partial_1^2 + \tilde D_2^2) = p_1^2 + \left(p_2 +eB x_1 \right)^2~,
}
$H\chi=m^2\chi$, \scalareom\ gives rise to a two dimensional scalar field $\varphi$ with mass $m$. The Hamiltonian \hamiltonianB\ is just that of the Landau problem of a particle in a magnetic field, whose spectrum is given by  
\eqn\shmspectrum{m_n^2 = (2n+1)|eB|~.}
Thus, turning on $B$ leads to a discrete spectrum of $1+1$ dimensional massive particles, as expected. 

As mentioned in the introduction, to preserve supersymmetry we must turn on in addition to the magnetic field also the $D$ component of the corresponding vector multiplet. The latter contributes to the scalar Lagrangian the term $eD|\phi|^2$, which shifts  \shmspectrum\ to
\eqn\shmspectrum{m_n^2 = (2n+1)|eB|-eD~.}
For the case $B=D$, for which the background fields preserve $(0,2)$ supersymmetry, we see that fields with $eB>0$ give rise to massless $1+1$ dimensional scalars, while those with $eB<0$ lead to a massive spectrum. Without loss of generality we can restrict to the case $B>0$ (otherwise, we can take $A_\mu, D, e$ to minus themselves), so that fields with positive $U(1)$ charge are the ones that give massless fields in two dimensions. As is familiar from the Landau problem, the spectrum \shmspectrum\ is in general degenerate. We will discuss this degeneracy below when we turn to the compact case. 

The above discussion can be repeated for spin $1/2$ fields. A four dimensional charged Weyl fermion in a magnetic field satisfies the wave equation 
\eqn\psieom{
i\left(\matrix{ -\partial_0 - \partial_3 & -\partial_1 + i \tilde D_2 \cr -\partial_1 - i \tilde D_2 & -\partial_0 + \partial_3  } \right)\left(\matrix{ \psi_- \cr \psi_+ } \right) = 0.
}
The top component of the spinor $(\psi_-)$ is a left-moving fermion in the two dimensions $(x^0,x^3)$; the bottom component, $\psi_+$, is right-moving. Squaring \psieom\ yields decoupled equations for $\psi_\pm$,
\eqn\psidc{
(-\partial_0^2 + \partial_3^2+\partial_1^2 + \tilde D_2^2 \mp i[\partial_1,\tilde  D_2])\psi_\pm=0~.}
Using the fact that $[\partial_1,\tilde  D_2] = [\partial_1, \partial_2 + ie Bx_1] = ie B$, \psidc\ is essentially identical to \hamiltonianB. So the right- and left-moving fermions have the spectrum
\eqn\psispectrum{
m^2_+= (2n+1)|eB| - eB, \qquad m^2_-: (2n+1)|eB|  + eB~.
}
Comparing to \shmspectrum\ we see that for $B=D$ the right-moving fermions align with the scalars, while the left-moving fermions do not, in agreement with the expectation from $(0,2)$ supersymmetry. 

To summarize, a four dimensional free massless chiral superfield $\Phi$ with $U(1)$ charge $e$ reduces in a constant $B=D>0$ background to 
a massless $(0,2)$ chiral superfield \phie\ for $e>0$, and to a massless Fermi superfield \lambdae\ for $e<0$. In both cases one also finds an infinite tower of massive two dimensional chiral and Fermi superfields with the spectrum \shmspectrum, \psispectrum. 


So far we have discussed the effects on the spectrum of charged fields of turning on a magnetic field in non-compact four dimensional spacetime. 
Since we are interested in turning on a magnetic field on $T^2$, we need to take the coordinates $(x^1, x^2)$ to be periodic, $x^i\sim x^i+2\pi R_i$.  The background gauge field $A_2 = Bx_1$ is not periodic on the torus; rather, it satisfies 
\eqn\peramu{A_2(x^1+2\pi R_1)=A_2(x^1)+2\pi R_1B=A_2(x^1)+\partial_2\Gamma~,}
with $\Gamma(x^2)=2\pi R_1Bx^2$ a gauge transformation parameter. 

A field $\phi$ of charge $e$ transforms under the gauge transformation generated by $\Gamma(x^2)$ as $\phi\to \exp(ie\Gamma)\phi$. Requiring that $\exp(ie\Gamma)$ is well defined on the torus leads to the Dirac quantization condition for the magnetic field
\eqn\dirquant{eB\AA\in 2\pi{\rm Z}}
where $\AA = 2\pi R_1  \times 2\pi R_2$ is the area of the torus. 

The eigenvalue problem for wavefunctions on the magnetized torus is now more complicated, due to the periodicity conditions. One finds (see \eg\ \CremadesWA) that the spectrum is still given by \shmspectrum, \psispectrum, and the degeneracy of states at a given level is 
\eqn\nnee{n_e=\frac{|e|B\AA}{2\pi}~.} 
Since the charges are proportional to the degeneracies with a universal proportionality constant, one can normalize them such that a field of charge $e$ has degeneracy $|e|$; we shall use this normalization in our discussion below. Thus, a field of charge $e>0$ gives $e$ massless $(0,2)$ chiral superfields, while one of charge $e<0$ gives $|e|$ massless Fermi superfields.  

\newsec{4d N=1 SQFT on a magnetized torus}

So far we discussed the effect of a magnetic field on free superfields in four dimensions; in this section we generalize to the interacting case. We take as the starting point an $N=1$ supersymmetric gauge theory with gauge group $G$ and chiral matter fields $\Phi_i$ in representations $r_i$ of the gauge group. There can also be a (gauge invariant) superpotential and other interactions, which we shall discuss later.

If we compactify such a theory on a two-torus without turning on a magnetic field, we find at low energies a two dimensional $(2,2)$ supersymmetric theory, which contains the $(2,2)$ chiral superfields $\Phi_i$ and a twisted chiral superfield $\Sigma$ in the adjoint representation describing the field strength of $G$.  In order to reduce the supersymmetry to $(0,2)$ we need to identify a suitable $U(1)$ global symmetry. We assign to the superfields $\Phi_i$ global charges $e_i$ and demand that the symmetry be non-anomalous (\ie\ conserved in the quantum theory). This leads to the constraints 
\eqn\anomfree{\eqalign{\sum_i e_i T(r_i)=&0~,\cr
\sum_i e_i^2 {\rm Tr} T^a(r_i)=&0~,} }
where $T^a(r)$ are the generators of $G$ in the representation $r$ and $T(r)$ is defined by ${\rm Tr}_r T^a T^b=T(r)\delta^{ab}$, with $a,b=1,\cdots, {\rm dim}\;G$. The first condition in \anomfree\ comes from the anomaly of one global and two gauge currents, while the second is the anomaly of two global and one gauge currents. We need to impose it since there is a non-zero source for the global $U(1)$. Note that we do not need to impose the vanishing of the anomaly of three global currents since $F\tilde F=0$ for it.

In general there may be many solutions to \anomfree, which correspond to different $U(1)$ subgroups of the global symmetry group of the model. We will comment on the dependence on the choice of $U(1)$ below. Note also that solutions to \anomfree\ include $U(1)$ factors of the gauge group. For those, one can show that the $B$ and $D$ fields that play a role in our construction are equivalent to a $B$ field and FI term for the dynamical $U(1)$. We will not discuss these cases in detail here, and will choose the $U(1)$ to be orthogonal to the gauge group. 

Four dimensional theories of the sort discussed above typically develop strong coupling in the UV or IR, when studied on $\IR^{3,1}$. If $T({\rm adj})>\sum_i T(r_i)/3$, the four dimensional theory is asymptotically free in the UV, and in general develops strong coupling in the IR below the dynamically generated scale $\Lambda$. One can then ask whether/how we can apply the results of the previous section to study the effects of the $B$ and $D$ fields on such a theory. As we saw, turning on the external fields leads to the appearance of a new energy scale associated with the Landau levels \shmspectrum. The other relevant scale is the Kaluza-Klein scale associated with the size of the two-torus. If these scales are much larger than $\Lambda$, we can use the results of the previous section to analyze the theory, since at the energies at which the $B$ field and compactification modify the dynamics, the four dimensional theory is still weakly coupled. 

If the hierarchy of scales is the other way around, naively we cannot use the results of the previous section, and have to search for a description of the four dimensional theory that is weakly coupled at the scales associated with the $B$ field and compactification. As usual, we may hope that since the two regimes are connected by a continuous deformation (the size of the torus), if there are no phase transitions as we vary the parameters, we nevertheless should find the right result by following the procedure of the previous section. We shall see examples of this later in the paper. 

Clearly, any non-trivial solution to \anomfree\ must involve some positive and negative $e_i$. As mentioned above, each field with $e_i>0$ gives $e_i$ $(0,2)$ massless chiral superfields $\phi_i$, while fields with $e_i<0$ give $|e_i|$ Fermi superfields. Fields with $e_i=0$, which include the gauge multiplet and any chiral superfields that are not charged under the $U(1)$, are not influenced by the background fields and can be treated by standard Kaluza-Klein methods. In particular, the gauge multiplet gives rise to the $(0,2)$ multiplets $\Upsilon, \Sigma$ as in \ttsigmad, while uncharged chiral superfields give chiral and Fermi superfields \ttphid\ in the appropriate representations of $G$. 

To find the two dimensional Lagrangian for these fields, we start with the four dimensional Lagrangian
 \eqn\indlll{\eqalign{&\LL =  \LL_V + \sum_{i}\LL_i ~,\cr
& \LL_V = -{1\over 4g^2}\left( F_{\mu\nu}^aF^{a\mu\nu} + 4i\bar\lambda^a \bar\sigma^\mu D_\mu\lambda^a -2 D^aD^a\right)~, \cr
&\LL_i = -D^\mu\bar\phi_i D_\mu\phi_i - i\bar\psi_i\bar\sigma^\mu D_\mu\psi_i + \bar\FF_i\FF_i+i\sqrt{2}(\bar\phi_i T^a\psi_i\lambda^a - \bar\lambda^a \bar\psi_i T^a\phi_i) + \bar\phi_i T^a\phi_i D^a~,
}}
where $\phi_i$, $\psi_i$ are components of the chiral superfields $\Phi_i$ and  $\lambda^a$ are the gauginos of $G$. We need to couple  \indlll\ to the background fields described above and reduce it on the torus, using the wavefunctions of the various fields. Although we are interested in low energy dynamics, some of the contributions of massive modes to \indlll\ need to be kept, since integrating them out may give terms in the Lagrangian of the massless modes that are relevant in the infrared. 

The Lagrangian of the vector superfield, $\LL_V$, gives rise to the standard $(2,2)$ Lagrangian \ttsigma, or equivalently the Lagrangian for the (0,2) gauge superfield $\Upsilon$ and chiral superfield in the adjoint representation,\foot{$\Sigma^{(0,2)}$ in \ttsigmad.} $\Sigma$, given in the first and second line of \saction. The bottom component of $\Sigma$ is a scalar field 
\eqn\sisi{\sigma=\frac{A_1+iA_2}{\sqrt2} }
that comes from components of the gauge field along the two-torus. 

Turning to the chiral superfields $\Phi_i$, we need to discuss separately the cases of fields with positive, negative and zero $U(1)$ charge. For fields with $e_i=0$, the compactification preserves $(2,2)$ SUSY and the Lagrangian is the usual one, reviewed in section 2. Fields with $e_i<0$ give rise to Fermi multiplets $\Lambda_i$, and their Lagrangians are given by the last line of \saction, \scomponent. The holomorphic functions $E_i$ \dlambda\ vanish in the massless sector, but receive non-zero contributions  \ttphie\ from massive modes. 

The reduction of fields with $e_i>0$ is more subtle. These fields give $(0,2)$ chiral superfields, which we shall also denote by $\Phi_i$, dropping the superscript $(0,2)$ in \ttphid. Before turning on the magnetic field, the Lagrangian of the scalars $\phi_i$ that are the bottom components of these superfields contains a potential proportional to $|\sigma|^2|\phi_i|^2$, which in $(0,2)$ language comes from the $|E_i|^2$ terms in \scomponent, with $E_i$ given by \ttphie. However, after turning on the magnetic field the $E_i$ must vanish in the light sector, for the same reason as in the $e_i<0$ case: the Fermi superfields associated with $\Phi_i$ are lifted, and there are no Fermi superfields with the right quantum numbers in the light sector  to give rise to a $|\sigma|^2|\phi_i|^2$ potential. 

To see how this happens, consider a massless four dimensional complex scalar field $\phi$ charged under a dynamical $U(1)$ gauge field $A_\mu$ and under a global $U(1)$ for which we turn on (equal) background $B$ and $D$ fields. The kinetic term for $\phi$, 
\eqn\lphi{\LL_\phi=-D_\mu\phi D^\mu\bar\phi+eD|\phi|^2+\cdots}
contains couplings to the dynamical and background $U(1)$ gauge fields $A_\mu$, $\tilde A_\mu$, $D_\mu\phi=(\partial_\mu+iA_\mu+ie\tilde A_\mu)\phi$. The terms with $\mu=1,2$, in particular, contain the coupling to the background $B$ field and the zero mode of the dynamical gauge field, $\sigma$ \sisi. Plugging the background gauge potential $\tilde A_2=Bx_1$ into \lphi\ and using \sisi, we see that the role of a non-zero $\sigma$ is to shift the location of the zero mode of $\phi$ in the $x^1$ plane. It has the same effect on the wavefunctions as a Wilson line for $\tilde A_\mu$, discussed \eg\ in \CremadesWA. 

For non-abelian gauge groups we do not expect the $|\sigma|^2\phi_i^2$ terms in the potential to completely disappear. The $(0,2)$ D-term potential includes terms of the form  $\bar\phi_i[\bar\sigma,\sigma]\phi_i$. They can be obtained by reducing the four dimensional action to two dimensions, taking into account the massive modes. 

To summarize, starting with a four dimensional $N=1$ SUSY gauge theory with gauge group $G$,  and chiral matter superfields $\Phi_i$ in representations $r_i$ of the gauge group, and compactifying it on a two-torus with equal background magnetic and $D$ fields for a global $U(1)$ symmetry under which $\Phi_i$ have charges $e_i \in {\rm Z} $, leads at low energies to a two dimensional $(0,2)$-supersymmetric gauge theory containing:

\item{(1)} A $G$ gauge superfield $\Upsilon$.
\item{(2)} An adjoint chiral superfield $\Sigma$.
\item{(3)} $e_i$ (0,2) chiral superfields $\Phi_i$ in the representation $r_i$, for fields with $e_i>0$.
\item{(4)} $|e_i|$ Fermi superfields $\Lambda_i$ in the representation $r_i$, for $e_i<0$.
\item{(5)} A chiral superfield $\Phi_i$ and Fermi superfield $\Lambda_i$ in the representation $r_i$, for $e_i=0$.

\noindent
In addition to the gauge interactions described in section 2, the fields above couple to the massive modes. These couplings are important for understanding some aspects of the dynamics.

The global symmetries of the resulting theory can be described from either the four dimensional or two dimensional point of view. Consider, for example, a $U(1)$ symmetry that assigns charges $q_i$ to the chiral superfields $\Phi_i$. In four dimensions, we have to impose the conditions\foot{Note that \anomfree\ is a special case of this, with $q_i=e_i$.} 
\eqn\globanom{\eqalign{\sum_i q_i T(r_i)=&0~,\cr
\sum_i e_i q_i {\rm Tr} T^a(r_i)=&0~,} }
coming from anomalies of one global and two gauge, and one global, one background and one gauge currents. The first of these is an inherently four dimensional constraint,\foot{Therefore, it can be lifted in the two dimensional infrared theory, if all the terms in the Lagrangian that violate symmetries that do not satisfy this constraint are irrelevant in the IR.} while the second has a natural two dimensional interpretation -- it is the two dimensional gauge anomaly of one global and one gauge current in the gauge theory with matter content (1) -- (5). 

An example of the above construction that will play a central role in our discussion below is supersymmetric QCD with gauge group $G=U(N_c)$ and $N_f$ flavors of chiral superfields in the fundamental representation of the gauge group, $Q^i$, $\tilde Q_i$, $i=1,2,\cdots, N_f$.  This theory has been extensively studied in the past, and has been found to exhibit a rich dynamical structure; see \eg\  \refs{\IntriligatorAU,\TerningBQ} for reviews. For $0<N_f<N_c$ it exhibits runaway behavior (no supersymmetric vacuum), while for $N_f\ge N_c$ it has non-trivial infrared behavior, and exhibits interesting dynamical phenomena such as quantum deformations of the classical moduli space, confinement, and Seiberg duality. 

The  global (non-R) symmetry group of this theory is 
\eqn\globsym{SU(N_f)\times SU(N_f)~.}
The two factors act by special unitary transformations on the $Q$'s and $\tilde Q$'s. For our construction we need to pick a $U(1)$ subgroup of \globsym, which can be done by assigning integer charges $e_i$ to $Q^i$ and $\tilde e_i$ to $\tilde Q_i$. The anomaly freedom constraints \anomfree\ imply that 
\eqn\anomsqcd{\eqalign{\sum_i e_i+\sum_i\tilde e_i=&0~,\cr
\sum_i e_i^2-\sum_i\tilde e_i^2=&0~.\cr}} 
The simplest solution to \anomsqcd\ is $e_i=e$, $\tilde e_i=-e$. However, this corresponds to picking the $U(1)$ to be the baryon number, which is gauged in our model. If we want the global $U(1)$ used for our construction to be orthogonal to the gauge group, we need to further require 
\eqn\stronganom{\sum_i e_i=\sum_i\tilde e_i=0~.}
An example of a solution to \stronganom, which exists for all even $N_f$, is to take $N_f/2$ of the $e_i$ to be equal to $+1$ and the rest equal to $-1$, and similarly for $\tilde e_i$. This breaks the symmetry \globsym\ to 
\eqn\globalsym{SU(N_f/2)^4\times U(1)~.}
The $N_f$ fundamentals $Q^i$ give rise to $N_f/2$ chiral superfields and $N_f/2$ Fermi superfields in the fundamental, and similarly for $\tilde Q$. Much of our discussion below will focus on this example.

More generally, we can take $N_+$ of the $e_i$ to be positive and $N_-=N_f-N_+$ to be negative, and similarly for $\tilde e$ (while imposing  \anomsqcd\ and \stronganom). The resulting theory has $\sum_{i \in N_+} e_i$ chiral multiplets and $\sum_{i \in N_-} e_i $ Fermi multiplets in the fundamental, and $\sum_{ j \in \tilde N_+} \tilde e_j$ chiral and $\sum_{ j \in \tilde N_-} \tilde e_j$ Fermi multiplets in the antifundamental. We will not study this more general case in detail, but will comment on it later.

$N=1$ SQCD in four dimensions exhibits Seiberg duality \SeibergPQ, which is the conjecture that its infrared limit is equivalent to that of a different theory, which has gauge group $U(N_f - N_c)$, $N_f$ chiral multiplets $q_i$ in the fundamental, $\tilde q^i$ in the antifundamental, and a meson $M^i_j$ which is a singlet of the gauge group, and is the dual of the gauge invariant composite field $Q^i\tilde Q_j$ in the electric theory. The magnetic theory further includes a superpotential coupling the magnetic meson $M$ to the magnetic quarks $q$, $\tilde q$,
\eqn\magneticspot{
\WW  =  M^i_j q_i\tilde q^j~.
}
Part of the statement of the duality is the identification of the global symmetry group of the electric theory, \globsym, with the corresponding symmetry in the magnetic theory. Hence, given a choice of charges $e_i$, $\tilde e_i$ in the electric theory, we can write down the corresponding charges in the magnetic one. The magnetic quarks  $q_i$ have charge $-e_i$, $\tilde q^i$ have charge $-\tilde e_i$, and $M^i_j$ have charge $e_i+\tilde e_j$. The spectrum of the low energy $(0,2)$ theory can then be read off the general analysis above. The superpotential \magneticspot\ gives rise at low energies to a $(0,2)$ superpotential, which will be discussed below.

\newsec{Brane construction}

$N=1$ SQCD has a natural embedding in string theory as the low energy effective theory on a system of $D$-branes and $NS5$-branes.\foot{We shall not provide a self-contained discussion of this system, instead referring the reader to the review \GiveonSR.} This description was found in the past to be useful for analyzing both supersymmetric and non-supersymmetric aspects of the dynamics of the theory, and it is natural to ask whether that is the case here as well. In this section we shall describe the gauge theory construction of the previous section in terms of branes.

The brane system whose infrared limit corresponds to $N=1$ SQCD is shown in figure 1. It contains three types of BPS branes in type IIA string theory: Neveu-Schwarz (NS) fivebranes, Dirichlet fourbranes and sixbranes. All branes are extended in the $3+1$ dimensions $x^\mu$, $\mu=0,1,2,3$. The $N_c$ $D4$-branes are further stretched in the $x^6$ direction between two differently oriented $NS5$-branes, which we shall refer to as $NS$ and $NS'$-branes. The former are stretched in $(x^4,x^5)$; the latter in $(x^8,x^9)$. The $N_f$ $D6$-branes are extended in the directions $(x^7,x^8,x^9)$, and are placed between the fivebranes. 

\bigskip

\ifig\loc{The brane system that realizes $N=1$ SQCD in type IIA string theory consists of $N_c$ $D4$-branes stretched in the $x^6$ direction between two differently oriented $NS5$-branes. $N_f$ $D6$-branes intersect the $D4$-branes at a particular $x^6$.}
{\epsfxsize2in\epsfbox{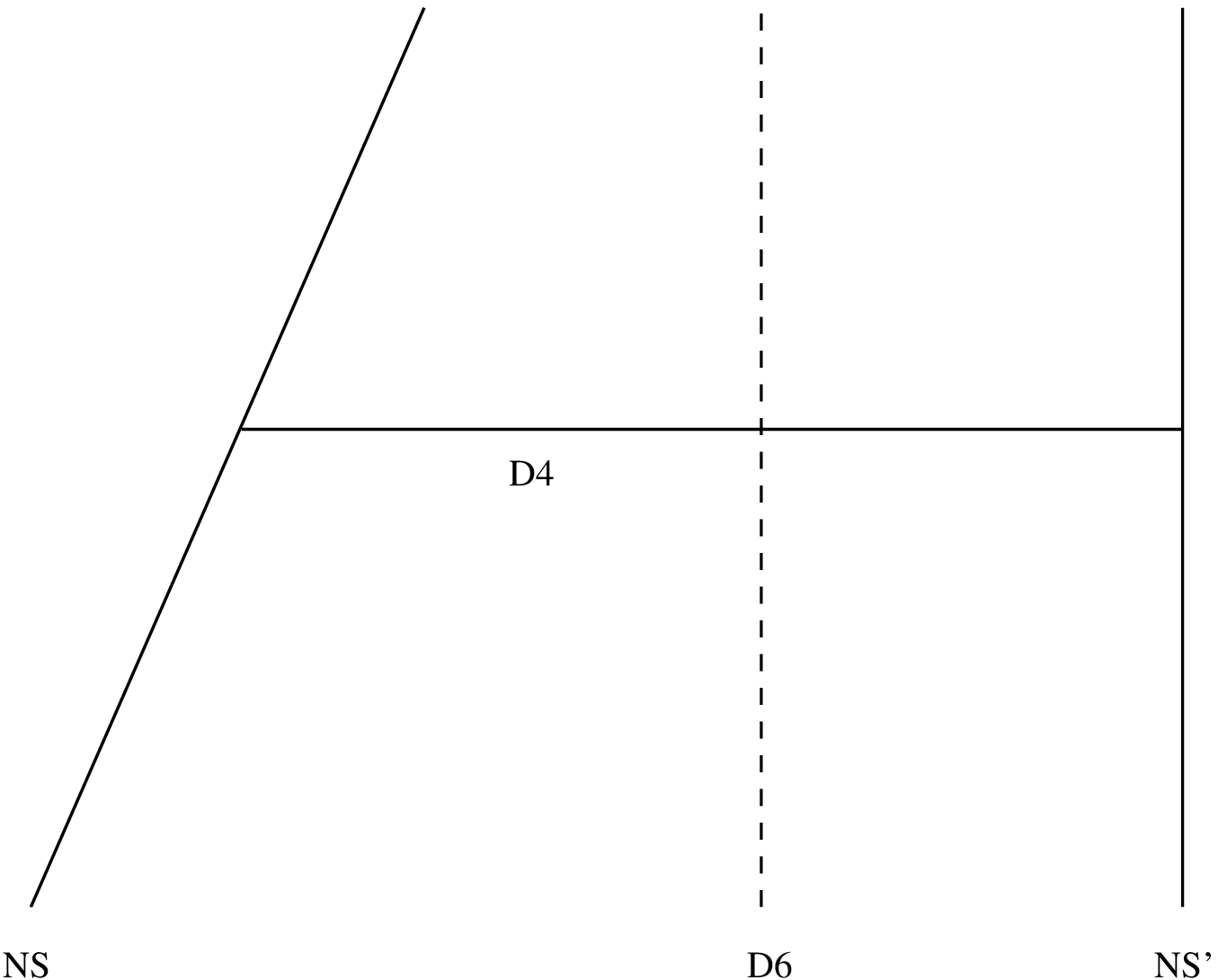}}

\bigskip

The configuration of figure 1 preserves four supercharges. Thus, the theory that lives at the intersection of the different branes is $N=1$ supersymmetric in the $3+1$ dimensions $x^\mu$; this theory is $N=1$ SQCD \ElitzurFH. The $U(N_c)$ vector superfield and corresponding SYM Lagrangian (the first line of \indlll) comes from taking the low energy limit of $4-4$ strings. For infinite $D4$-branes this gives five dimensional SYM with sixteen supercharges; however here one of the worldvolume directions $(x^6)$ is a line segment, so the low energy theory is $3+1$ dimensional. The boundary conditions at the ends of the line segment, where the fourbranes end on fivebranes, give mass to all fields other than the $N=1$ SYM ones. 

The $N_f$ fundamental chiral superfields $Q^i$, $\tilde Q_i$ come from $4-6$ strings which are localized in $x^6$. The full global symmetry \globsym\ is not manifest in figure 1. The diagonal $SU(N_f)$ is visible as the gauge symmetry on the flavor branes\foot{The gauge symmetry on the sixbranes is in fact $U(N_f)$, but the $U(1)$ acts on the low energy theory in the same way as the $U(1)$ factor in $U(N_c)$, and so is not an independent symmetry.} (the $D6$-branes). The full symmetry \globsym\ is from the point of view of figure 1 an accidental symmetry of the low energy theory. One can make it manifest by moving all the $D6$-branes in the $x^6$ direction to the location of the $NS'$-brane \BrodieSZ. Then the $D6$-branes are split into two disconnected components (with $x^7\ge0$ and $x^7\le 0$ respectively) by the fivebrane, and one can perform separate $SU(N_f)$ transformations on the two components. 

To implement our construction, we would like to compactify $(x^1,x^2)$ on a torus, pick a $U(1)$ inside the global symmetry group \globsym, and turn on the magnetic field on the torus and $D$ field for it. As mentioned above, in the brane construction it is slightly simpler to deal with the diagonal $SU(N_f)$;  therefore, we shall take the $U(1)$ to be a subgroup of this $SU(N_f)$. It is possible to generalize the discussion to other $U(1)$ symmetries, by placing the $D6$-branes of figure 1 at the location of the $NS'$-brane and using the results of \BrodieSZ.

In terms of the branes, the procedure of turning on the external fields discussed above is described as follows. Each of the $N_f$ $D6$-branes gives rise to one hypermultiplet, $Q^i$, $\tilde Q_i$, which is charged under the $U(1)$ gauge field living on the sixbrane. We turn on a magnetic field $B_i$ for this $U(1)$ field, and accompany it by a suitable rotation of the $D6$-brane from the $x^7$ to the $x^6$ direction. The latter is the brane analog of the D-term in the low energy gauge theory, and it preserves SUSY if we tune the rotation angle to correspond to the magnetic field that we turned on. This can be shown directly, but we shall see it in a slightly different language later. The requirement that the $B$ field lies in the diagonal $SU(N_f)$ (\ie\ is orthogonal to the gauge group, as in \stronganom) is in this language $\sum_i B_i=0$. The resulting configuration in the $(x^6, x^7)$ plane is exhibited in figure 2. The $N_f$ $D6$-branes are rotated from the $x^7$ axis by angles $\theta_i$ that are determined by the magnetic fields $B_i$, $\tan\theta_i\propto B_i$. This corresponds in the field theory to giving to the $i$'th flavor charge $e_i$ proportional to $B_i$ for $Q^i$ and $\tilde e_i=-e_i$ for $\tilde Q_i$.  

\bigskip

\ifig\loc{Turning on the $D$ field gives rise to a configuration in which the $D6$-branes are rotated in the $(x^6, x^7)$ plane.}
{\epsfxsize2in\epsfbox{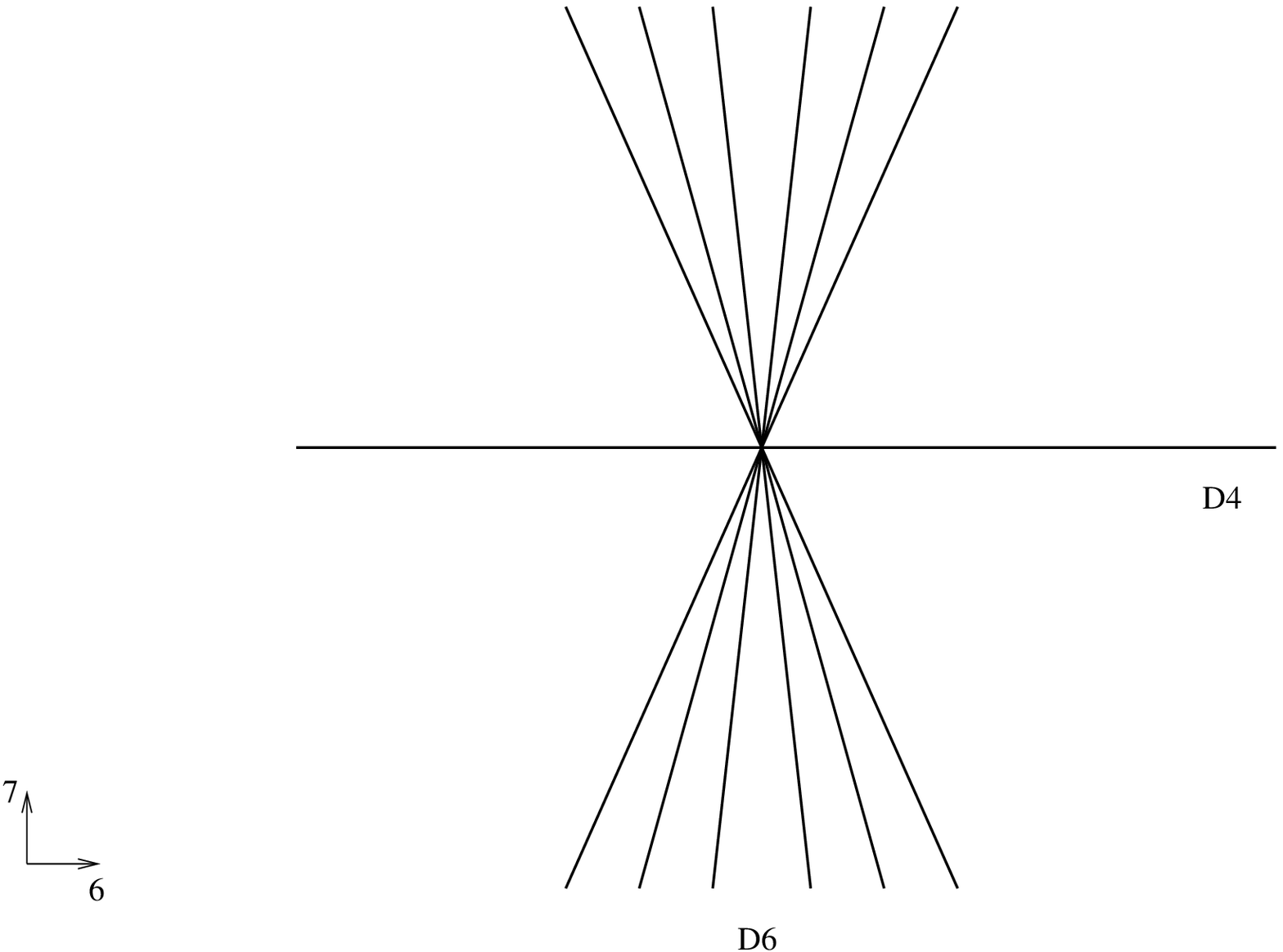}}

\bigskip

Since we shall be discussing below the consequences of Seiberg duality in the compactified theory with the background $B$ and $D$ fields, it is useful to review the generalization of the above discussion to the magnetic theory. The brane configuration corresponding to the Seiberg dual theory \refs{\ElitzurFH,\GiveonSR} is depicted in figure 3.  

The $N_f-N_c$ color $D4$-branes connecting the $NS5$-branes give rise to the magnetic gauge group $U(N_f-N_c)$. The $N_f$ flavor $D4$-branes stretched between the $NS'$-brane and $D6$-branes give the magnetic meson $M$. The magnetic quarks $q$, $\tilde q$ come from strings stretched  between the color and flavor branes, and are thus localized near the $NS'$-brane. 

\bigskip

\ifig\loc{The brane system that realizes the Seiberg dual of $N=1$ SQCD includes $N_f-N_c$ color $D4$-branes connecting two $NS5$-branes, and $N_f$ flavor $D4$-branes, each connecting the $NS'$-brane to one of $N_f$ $D6$-branes.}
{\epsfxsize2in\epsfbox{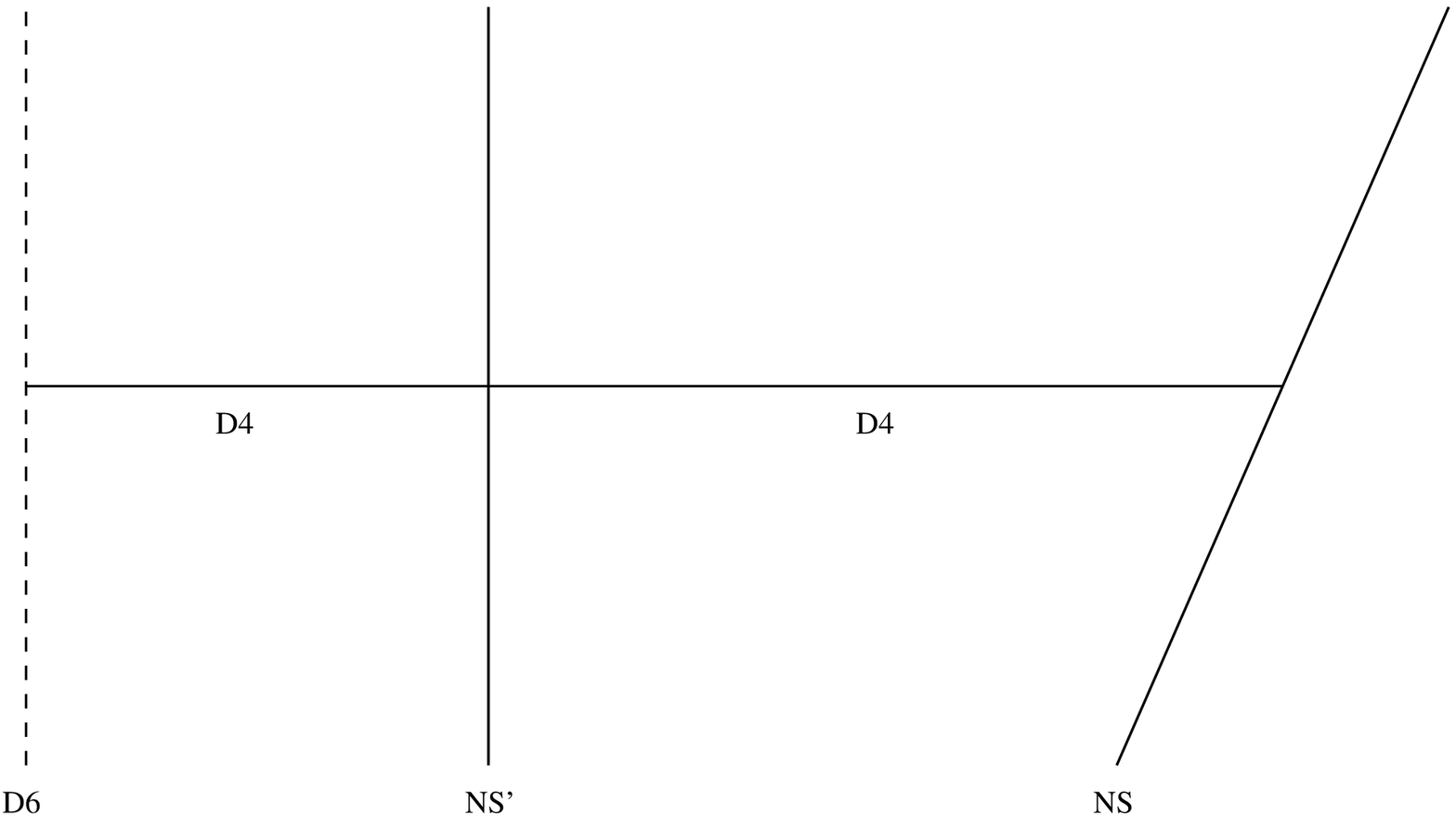}}

\bigskip

Turning on the $B$ field on the $D6$-branes and rotating them in the $(67)$ plane by a suitable amount leads to the configuration of figure 4. The flavor $D4$-branes are now rotated in the $(67)$ plane by the same angle as the corresponding $D6$-branes. Supersymmetry seems to be superficially violated, but is restored by a non-zero value of the magnetic field on the flavor $D4$-branes.

\bigskip

\ifig\loc{The effect of the $D$ field on the magnetic brane system.}
{\epsfxsize2in\epsfbox{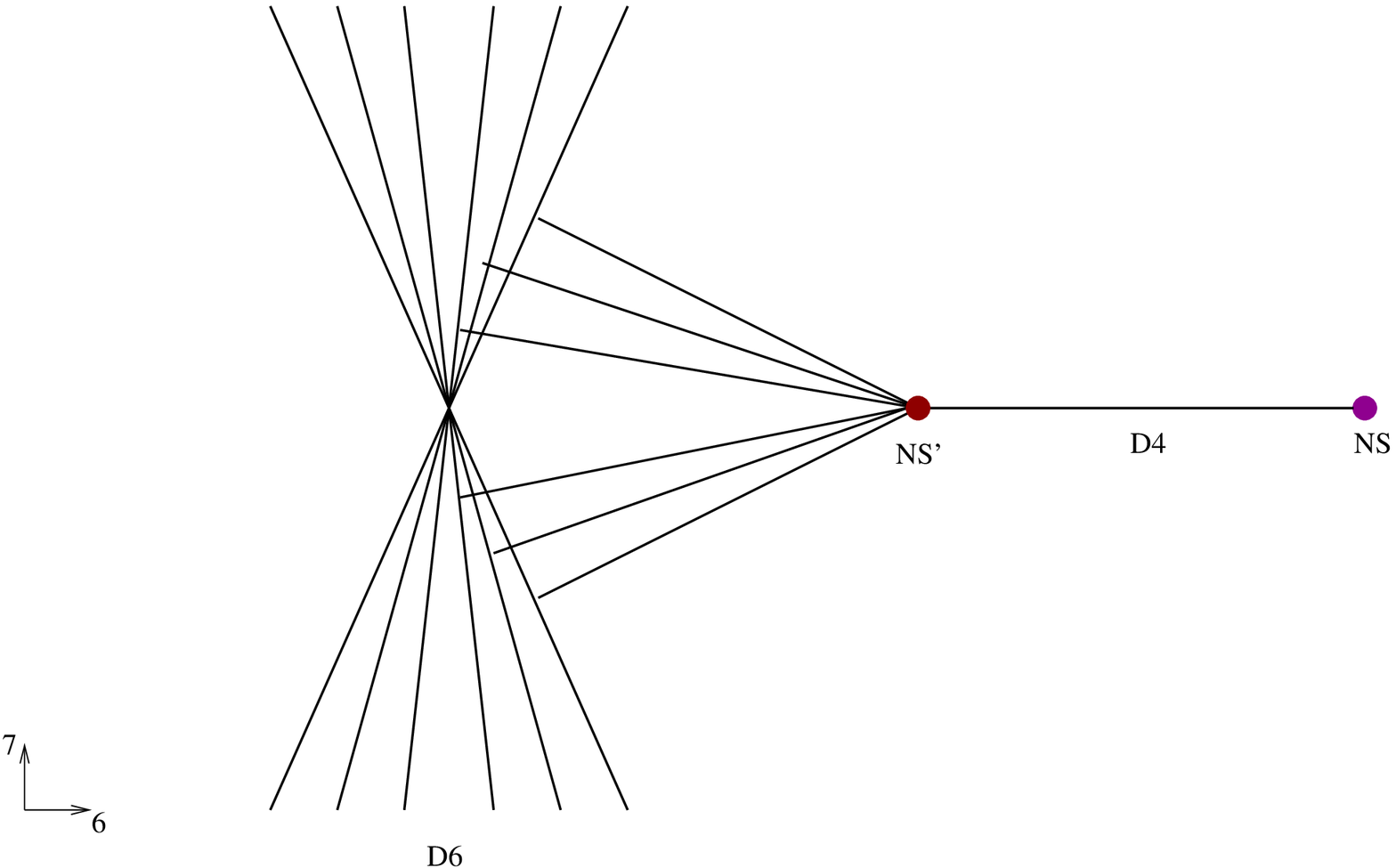}}

\bigskip

In the brane systems described in this section, the $D$ field deformation is described geometrically, but the magnetic field on the torus is not. It would be nice to geometrize the $B$ field as well. To this end we next turn to a description of our system as a compactification of the three dimensional theory obtained by first compactifying $N=1$ SQCD on a circle.

\newsec{Three dimensional description}

Three dimensional $N=2$ SQCD has a brane description which is formally obtained by T-dualizing the system of figure 1 in, say, the $x^2$ direction \GiveonSR. This does not do anything to the $NS5$-branes, but turns the $D4$-branes into $D3$-branes stretched in $(0136)$, and the $D6$-branes into $D5$-branes stretched in $(013789)$, in type IIB string theory. Since we want to think of the theory as a compactification of four dimensional $N=1$ SQCD on a finite circle, we keep the size of the $x^2$ circle finite. 

\bigskip

\ifig\loc{Quantization of the magnetic field on the two-torus corresponds in the IIB language to quantization of the angle that the $D5$-brane makes with the $x^1$ axis. Plotted is a $D5$-brane with $e=2$ in the double covering space of the torus (parallel dashed lines are identified). As the $D5$-brane wraps the $x^1$ circle once, it wraps the $x^2$ circle twice.
}
{\epsfxsize2in\epsfbox{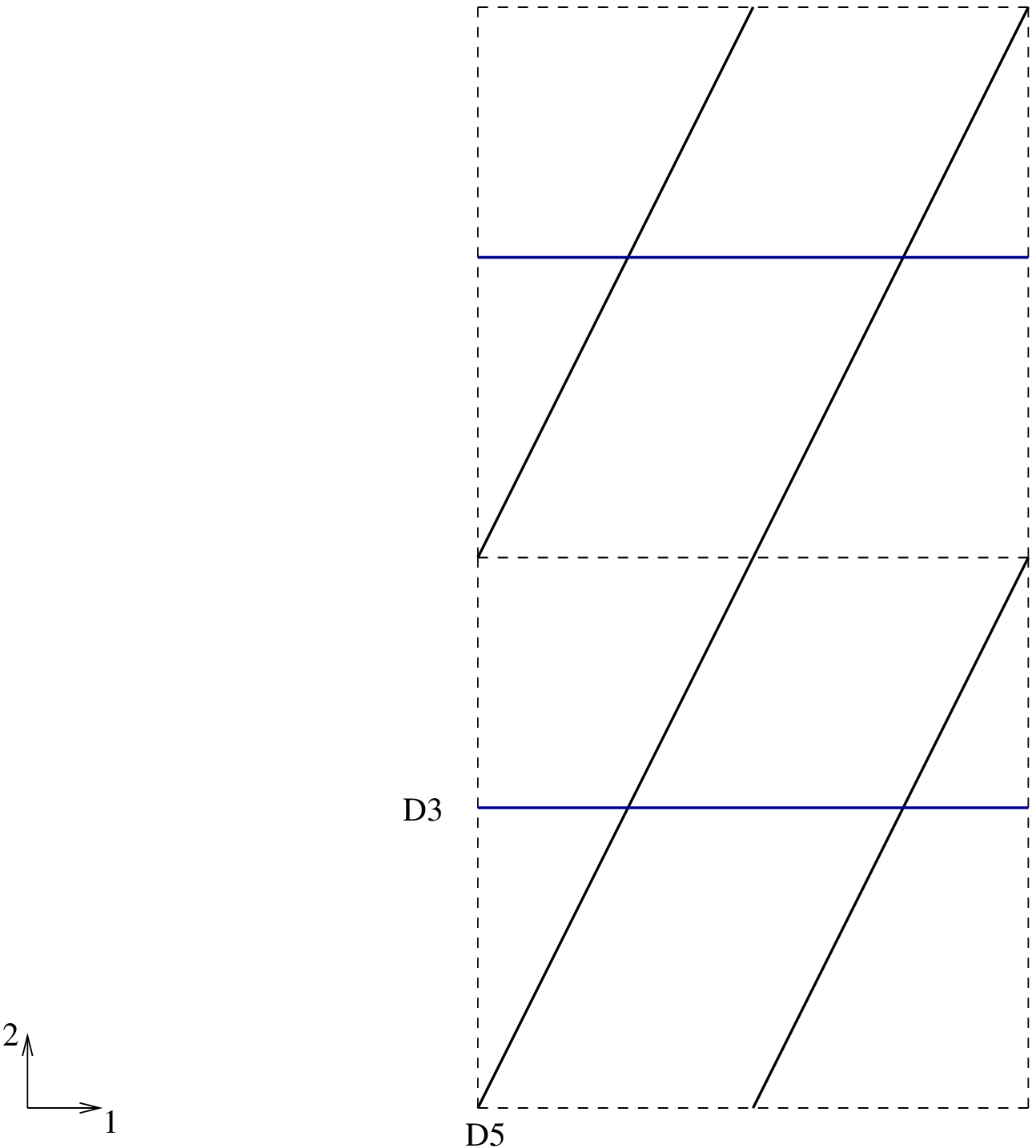}}

\bigskip

As before, we want to also compactify $x^1$ on a circle and turn on the background fields $B$ and $D$ described above. In the four dimensional description, the $B$ field corresponds to a vector potential $A_2=Bx^1$ for a global $U(1)$, which is realized in the brane construction as the $U(1)$ gauge field on a $D6$-brane. In three dimensions, $A_2$ becomes a scalar field in the vector multiplet on a $D5$-brane. Turning on an expectation value for it corresponds to rotating the $D5$-brane in the $(x^1, x^2)$ plane. The quantization of the $B$ field is manifest in the three dimensional description: it is due to the fact that as the $D5$-brane wraps the $x^1$ circle once, it has to return to its original position on the torus, and thus must wrap the $x^2$ circle an integer number of times (see figure 5).  As is clear from the figure, rotating the $D5$-brane in this way localizes the matter coming from $3-5$ strings on the $x^1$ circle, as is expected from the four dimensional perspective, where this is due to the magnetic field. The degeneracy \nnee\ comes in this language from the fact that a $D5$-brane that wraps $e$ times around the $x^2$ circle as it goes once around $x^1$ intersects the $D3$-branes at $e$ points on the $x^1$ circle.

In the low energy field theory, moving a $D5$-brane in $x^2$ corresponds to giving equal and opposite real masses to the corresponding chiral superfields $Q$, $\tilde Q$. Rotating it in the $(x^1,x^2)$ plane thus corresponds to a real mass that depends linearly on $x^1$, which breaks Lorentz symmetry to $SO(1,1)$ and localizes these fields at the minimum of the resulting potential. 

Turning on the $D$ field again corresponds to rotating the $D5$-brane in the $(x^6,x^7)$ plane, as in figure 2. The fact that the configuration preserves supersymmetry can be seen as in other cases involving rotated branes \BerkoozKM. Defining $z_1=x^1+ix^7$ and $z_2=x^2+ix^6$, the $D5$-branes are originally located at $z_2=0$ (say), and together with the other branes in the configuration preserve $N=2$ supersymmetry in the three dimensions $(x^0,x^1,x^3)$.  Rotating a $D5$-brane by a general angle $\theta$, 
\eqn\rotatezz{z\to \Omega(\theta) z~,}
with $\Omega(\theta)$ the standard $2\times 2$ rotation matrix, preserves two of the four supercharges, which form a $(0,2)$ superalgebra in the $1+1$ dimensions $(x^0, x^3)$.  

We can now combine all of the above elements to describe the models of section 4 in terms of branes. Consider \eg\ the model in which we give $N_f/2$ of the $Q$'s ($\tilde Q$'s) charge $e=+1 (-1)$, and to the other $N_f/2$ the opposite charge. The corresponding brane configuration is depicted in figure 6. Each intersection of the $N_c$ $D3$-branes with one of the $N_f$ rotated $D5$-branes supports either a $(0,2)$ chiral superfield coming from $Q$ and a Fermi superfield from $\tilde Q$, or the other way around, depending on the sign of the rotation angle. The $\Sigma$ chiral superfield comes from the low lying modes living on the $D3$-branes, as does the gauge superfield $\Upsilon$. In the next section we shall study the dynamics of these fields.

\bigskip

\ifig\loc{Two views of the brane configuration describing compact three dimensional $N=2$ SQCD in a background global $U(1)$ under which half of the flavors have charge $+1$ and half $-1$. Panel (a) describes the configuration in the $(12)$ plane; the flavor branes wind once around the $x^2$ circle as they wind once around the $x^1$ one. Panel (b) shows the configuration in the $(67)$ plane; turning on the $D$ field for the flavor symmetry corresponds to a rotation of the flavor branes from $x^7$ to $x^6$.
}
{\epsfxsize2.3in\epsfbox{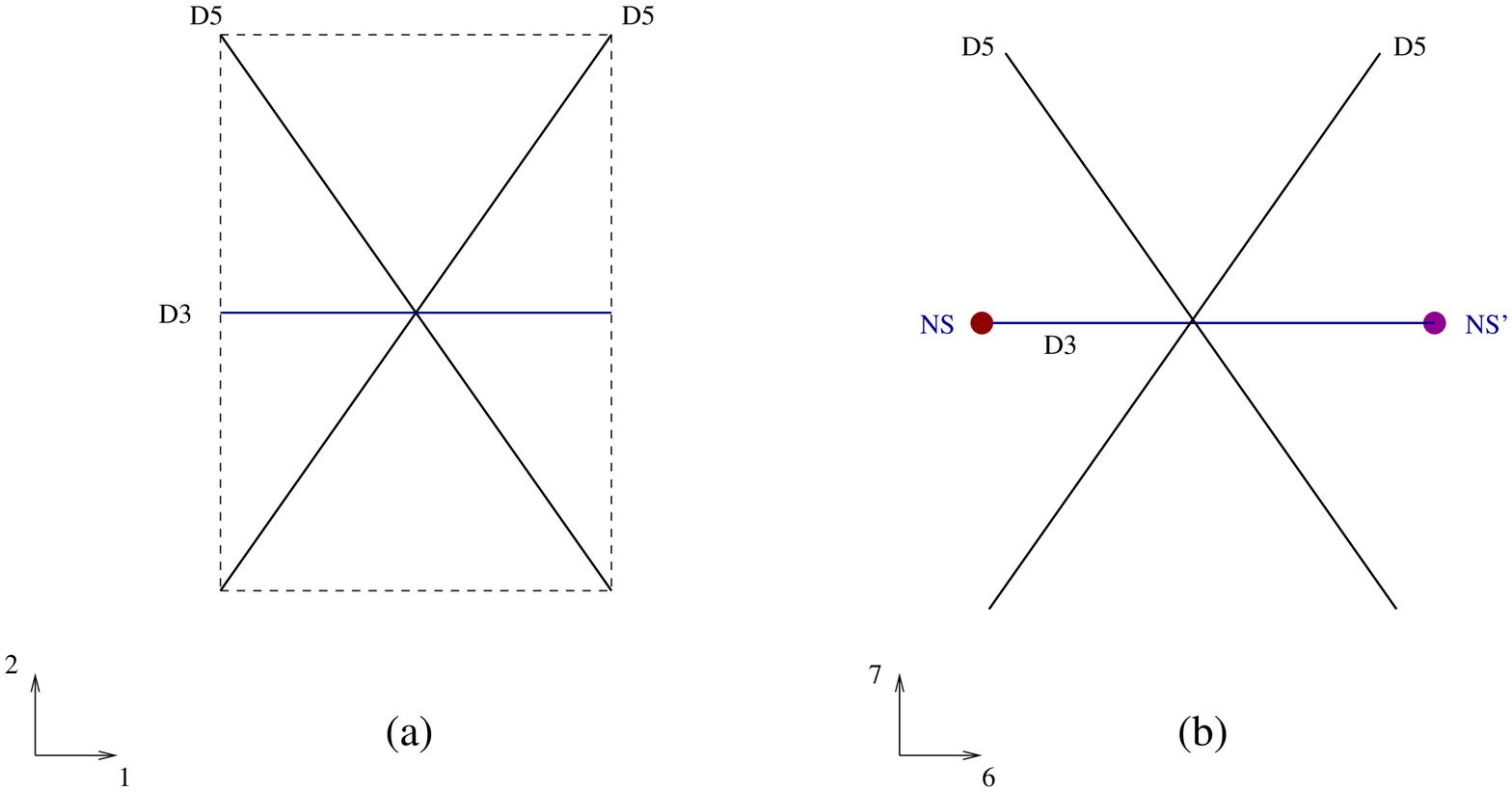}}

\bigskip

Note, incidentally, that if we were to remove the constraint \stronganom\ that the $U(1)$ which we use for the construction is not part of the gauge group, we could consider taking the charges to be $e_i=e$, $\tilde e_i=-e$ for all $i=1,\cdots, N_f$. This corresponds to turning on a magnetic field and $D$ field for the $U(1)$ factor in the gauge group, $U(1)_B$. In the brane picture, this would correspond to rotating all $N_f$ $D5$-branes by the same angle, leading to the brane configuration of figure 7 (a). 

\bigskip

\ifig\loc{Turning on $B$ and $D$ fields for a global symmetry that acts on the dynamical fields in the same way as the $U(1)$ part of the gauge group leads to the brane configuration (a), which is equivalent to the configuration (b) obtained by turning on an FI term and scalar field $\phi_2(x^1)$ for that $U(1)$.}
{\epsfxsize2.3in\epsfbox{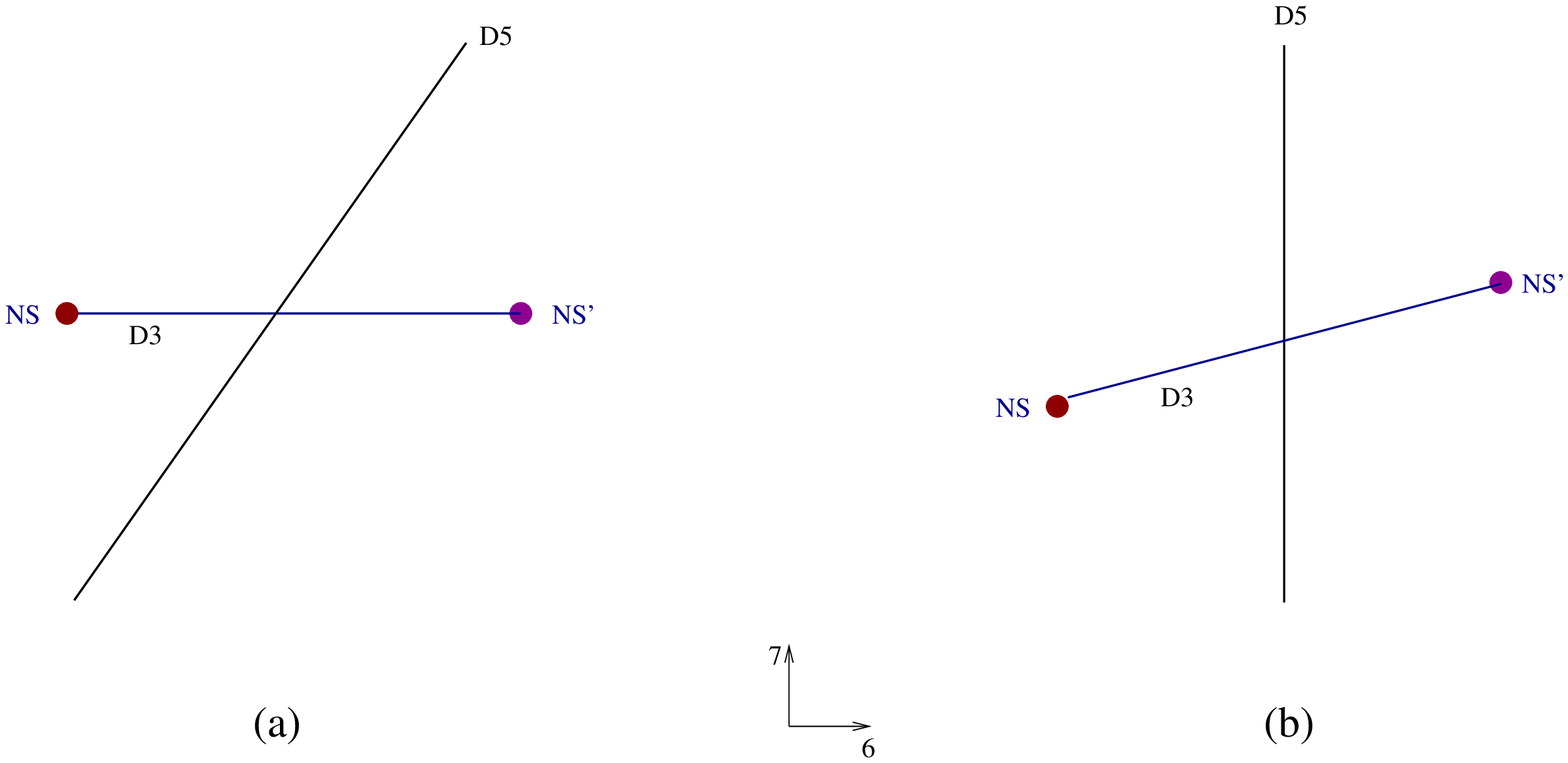}}

\bigskip

An equivalent brane configuration can be obtained by turning on an FI term for the $U(1)$ factor in the gauge group. Indeed, the latter corresponds to the relative displacement of the two $NS5$-branes in $x^7$ \GiveonSR. Naively, this deformation breaks supersymmetry; however there are two ways to restore it (classically). If the number of flavors is large enough, one can break the gauge group and go into the Higgs branch, while maintaining $2+1$ dimensional Poincare symmetry. Another option, which is available for all $N_f$, is to turn on the scalar field $\phi_2$ in the $U(1)$ vector multiplet, $\phi_2\propto x^1$, breaking the $2+1$ dimensional Lorentz symmetry down to $1+1$ dimensions. The resulting brane configuration is plotted in figure 7 (b). Clearly, figures 7 (a) and (b) are related by an overall rotation in the $(67)$ plane, and lead to equivalent physics. This is a quick way to see that rotating the flavor branes in the $(67)$ plane corresponds to turning on a D-term for a $U(1)$ symmetry.  Note that the above discussion also implies that turning on an FI term for a $U(1)$ gauge symmetry in the higher dimensional theory does not correspond to turning on the FI term \fiterm\ in the low energy two dimensional $(0,2)$ theory.

\newsec{Comments on the quantum theory}

In this section we  discuss some aspects of the low energy dynamics of the systems described in the previous sections. We will focus on the choice of charges depicted in figure 6, but it is easy to generalize to other cases.

The light spectrum of the resulting model is summarized in the following table, where we indicated the transformation properties of the superfields under the $SU(N_f/2)$ factors in \globalsym, as well as under the global $U(1)$ used in our construction, which is denoted by $U(1)_e$ in the table. The gauge transformation properties have been suppressed in the table. The first (last) two fields in the table transform in the (anti) fundamental representation of the $U(N_c)$ gauge group. 

\bigskip
\vbox{
$$\vbox{\offinterlineskip
\hrule height 1.1pt
\halign{&\vrule width 1.1pt#
&\strut\quad#\hfil\quad&
\vrule width 1.1pt#
&\strut\quad#\hfil\quad&
\vrule width 1.1pt#
&\strut\quad#\hfil\quad&
\vrule width 1.1pt#
&\strut\quad#\hfil\quad&
\vrule width 1.1pt#
&\strut\quad#\hfil\quad&
\vrule width 1.1pt#
&\strut\quad#\hfil\quad&
\vrule width 1.1pt#\cr
height3pt
&\omit&
&\omit&
&\omit&
&\omit&
&\omit&
&\omit&
\cr
&\hfil field&
&\hfil $SU(N_f/2)_1$&
&\hfil $SU(N_f/2)_2$&
&\hfil $SU(N_f/2)_3$&
&\hfil $SU(N_f/2)_4$&
&\hfil $U(1)_e$&
\cr
height3pt
&\omit&
&\omit&
&\omit&
&\omit&
&\omit&
&\omit&
\cr
\noalign{\hrule height 1.1pt}
height3pt
&\omit&
&\omit&
&\omit&
&\omit&
&\omit&
&\omit&
\cr
&\hfil $Q^1$&
&\hfil $N_f/2$&
&\hfil $1$&
&\hfil $1$&
&\hfil $1$&
&\hfil $+1$&
\cr
height3pt
&\omit&
&\omit&
&\omit&
&\omit&
&\omit&
&\omit&
\cr
\noalign{\hrule}
height3pt
&\omit&
&\omit&
&\omit&
&\omit&
&\omit&
&\omit&
\cr
&\hfil $\Lambda^2$&
&\hfil $1$&
&\hfil $N_f/2$&
&\hfil $1$&
&\hfil $1$&
&\hfil $-1$&
\cr
\noalign{\hrule}
height3pt
&\omit&
&\omit&
&\omit&
&\omit&
&\omit&
&\omit&
\cr
&\hfil $\tilde\Lambda_1$&
&\hfil $1$&
&\hfil $1$&
&\hfil $\bar{N_f/2}$&
&\hfil $1$&
&\hfil $-1$&
\cr
\noalign{\hrule}
height3pt
&\omit&
&\omit&
&\omit&
&\omit&
&\omit&
&\omit&
\cr
&\hfil $\tilde Q_2$&
&\hfil $1$&
&\hfil $1$&
&\hfil $1$&
&\hfil $\bar{N_f/2}$&
&\hfil $+1$&
\cr
height3pt
&\omit&
&\omit&
&\omit&
&\omit&
&\omit&
&\omit&
\cr
}\hrule height 1.1pt
}
$$
}
\centerline{\sl Table 1: The quantum numbers of the light states of the electric model.} 

In addition to the fields in Table 1, there is the $(0,2)$ chiral superfield $\Sigma$ and the field strength $\Upsilon$.  Both transform in the adjoint of $U(N_c)$ and are singlets under all the symmetries listed in Table 1. 
As explained in section 4, the $E$ functions for the Fermi superfields in Table 1 vanish in the light sector, and in particular there is no potential of the form $|\sigma|^2|\Phi|^2$ coupling the fundamentals $(\Phi=Q,\tilde Q)$ and adjoint. This is easy to see from the brane perspective. Before turning on the background $B$ and $D$ fields, this potential could be understood as follows. The imaginary part of $\sigma$ can be thought of as parametrizing the location of the $D3$-branes in the $x^2$ direction.\foot{On the Coulomb branch, the $U(N_c)$ gauge symmetry is broken to $U(1)^{N_c}$. The real part of $\sigma$ can be thought of as the dual of the unbroken gauge field.} For generic positions of the threebranes, strings stretched between them and the fivebranes have a minimal length proportional to the $D3-D5$ separation in the $x^2$ direction. This gives rise to a mass for the chiral superfields which is captured by the $|\sigma|^2|\Phi|^2$ potential. After turning on the background fields, it is no longer true that changing $\sigma$ gives a mass to the chiral superfields. As is clear from figure 6 (a), all it does is change their location in $x^1$, an effect that we saw in the discussion of the gauge theory Lagrangian \lphi.

One of our goals in this section is to compare the two dimensional theory obtained from SQCD to the one obtained from its Seiberg dual. To facilitate the comparison, we present in Table 2 the two dimensional spectrum and quantum numbers of the charged superfields in the magnetic theory. 

\bigskip
\vbox{
$$\vbox{\offinterlineskip
\hrule height 1.1pt
\halign{&\vrule width 1.1pt#
&\strut\quad#\hfil\quad&
\vrule width 1.1pt#
&\strut\quad#\hfil\quad&
\vrule width 1.1pt#
&\strut\quad#\hfil\quad&
\vrule width 1.1pt#
&\strut\quad#\hfil\quad&
\vrule width 1.1pt#
&\strut\quad#\hfil\quad&
\vrule width 1.1pt#
&\strut\quad#\hfil\quad&
\vrule width 1.1pt#\cr
height3pt
&\omit&
&\omit&
&\omit&
&\omit&
&\omit&
&\omit&
\cr
&\hfil field&
&\hfil $SU(N_f/2)_1$&
&\hfil $SU(N_f/2)_2$&
&\hfil $SU(N_f/2)_3$&
&\hfil $SU(N_f/2)_4$&
&\hfil $U(1)_e$&
\cr
height3pt
&\omit&
&\omit&
&\omit&
&\omit&
&\omit&
&\omit&
\cr
\noalign{\hrule height 1.1pt}
height3pt
&\omit&
&\omit&
&\omit&
&\omit&
&\omit&
&\omit&
\cr
&\hfil $\lambda_1$&
&\hfil $\bar{N_f/2}$&
&\hfil $1$&
&\hfil $1$&
&\hfil $1$&
&\hfil $-1$&
\cr
height3pt
&\omit&
&\omit&
&\omit&
&\omit&
&\omit&
&\omit&
\cr
\noalign{\hrule}
height3pt
&\omit&
&\omit&
&\omit&
&\omit&
&\omit&
&\omit&
\cr
&\hfil $q_2$&
&\hfil $1$&
&\hfil $\bar{N_f/2}$&
&\hfil $1$&
&\hfil $1$&
&\hfil $+1$&
\cr
\noalign{\hrule}
height3pt
&\omit&
&\omit&
&\omit&
&\omit&
&\omit&
&\omit&
\cr
&\hfil $\tilde q^1$&
&\hfil $1$&
&\hfil $1$&
&\hfil $N_f/2$&
&\hfil $1$&
&\hfil $+1$&
\cr
\noalign{\hrule}
height3pt
&\omit&
&\omit&
&\omit&
&\omit&
&\omit&
&\omit&
\cr
&\hfil $\tilde \lambda^2$&
&\hfil $1$&
&\hfil $1$&
&\hfil $1$&
&\hfil $N_f/2$&
&\hfil $-1$&
\cr
\noalign{\hrule}
height3pt
&\omit&
&\omit&
&\omit&
&\omit&
&\omit&
&\omit&
\cr
&\hfil $M^1_1,\Lambda^1_1$&
&\hfil $N_f/2$&
&\hfil $1$&
&\hfil $\bar{N_f/2}$&
&\hfil $1$&
&\hfil $0$&
\cr
\noalign{\hrule}
height3pt
&\omit&
&\omit&
&\omit&
&\omit&
&\omit&
&\omit&
\cr
&\hfil $M^2_2,\Lambda^2_2$&
&\hfil $1$&
&\hfil $N_f/2$&
&\hfil $1$&
&\hfil $\bar{N_f/2}$&
&\hfil $0$&
\cr
height3pt
&\omit&
&\omit&
&\omit&
&\omit&
&\omit&
&\omit&
\cr
\noalign{\hrule}
height3pt
&\omit&
&\omit&
&\omit&
&\omit&
&\omit&
&\omit&
\cr
&\hfil $M^1_2 (\times 2)$&
&\hfil $N_f/2$&
&\hfil $1$&
&\hfil $1$&
&\hfil $\bar{N_f/2}$&
&\hfil $+2$&
\cr
height3pt
&\omit&
&\omit&
&\omit&
&\omit&
&\omit&
&\omit&
\cr
\noalign{\hrule}
height3pt
&\omit&
&\omit&
&\omit&
&\omit&
&\omit&
&\omit&
\cr
&\hfil $\Lambda^2_1(\times 2)$&
&\hfil $1$&
&\hfil $N_f/2$&
&\hfil $\bar{N_f/2}$&
&\hfil $1$&
&\hfil $-2$&
\cr
height3pt
&\omit&
&\omit&
&\omit&
&\omit&
&\omit&
&\omit&
\cr
}\hrule height 1.1pt
}
$$
}
\centerline{\sl Table 2: The quantum numbers of the light states of the magnetic model.} 

\bigskip

\noindent
Here $\lambda_1$, $\tilde\lambda^2$ are Fermi superfields obtained from the four dimensional chiral superfields $q_1$, $\tilde q^2$, which have $U(1)_e$ charge $-1$, using the construction of section 3. Similarly, $q_2$ and $\tilde q^1$, which have charge $+1$, give rise to chiral superfields. The four dimensional singlet meson fields $M^i_i$, $i=1,2$, have $U(1)_e$ charge $0$ and thus give upon reduction a KK tower of chiral and Fermi superfields, the lowest of which are massless. The off-diagonal meson fields $M^1_2$ and $\Lambda^2_1$ have charges $+2$ and $-2$, respectively, and give chiral and Fermi superfields with degeneracy two. This degeneracy is due to the fact that the flavor $D3$-branes in the brane configuration dual to that of figure 6 intersect twice on the torus. Each intersection supports one copy of the above chiral and Fermi superfields. The magnetic superpotential \magneticspot\ gives rise in two dimensions to an effective $(0,2)$ superpotential of the schematic form 
\eqn\zerotwospot{\WW=M^1_1\lambda_1\tilde q^1+M_2^2q_2\tilde\lambda^2+\Lambda^2_1q_2\tilde q^1\,.}

We would like to understand the low energy dynamics of the electric and magnetic theories described above. Starting with the electric theory, the first issue we would like to discuss is the fate of the Coulomb branch. The D-term potential ${\rm Tr}\ [\sigma,\bar\sigma]^2$  can be used as usual to diagonalize the adjoint scalar field $\sigma$. The Coulomb branch is labeled by the eigenvalues of this matrix. From the brane perspective it corresponds to displacing the color branes in the $x^2$ direction, as depicted in figure 8 (a).

\bigskip

\ifig\loc{(a) At a generic point on the Coulomb branch, the $N_c$ $D3$-branes are separated in $x^2$. (b) At special points on the Coulomb branch, the $Q$'s and $\tilde Q$'s are localized at the same points in $x^1$.}
{\epsfxsize2.3in\epsfbox{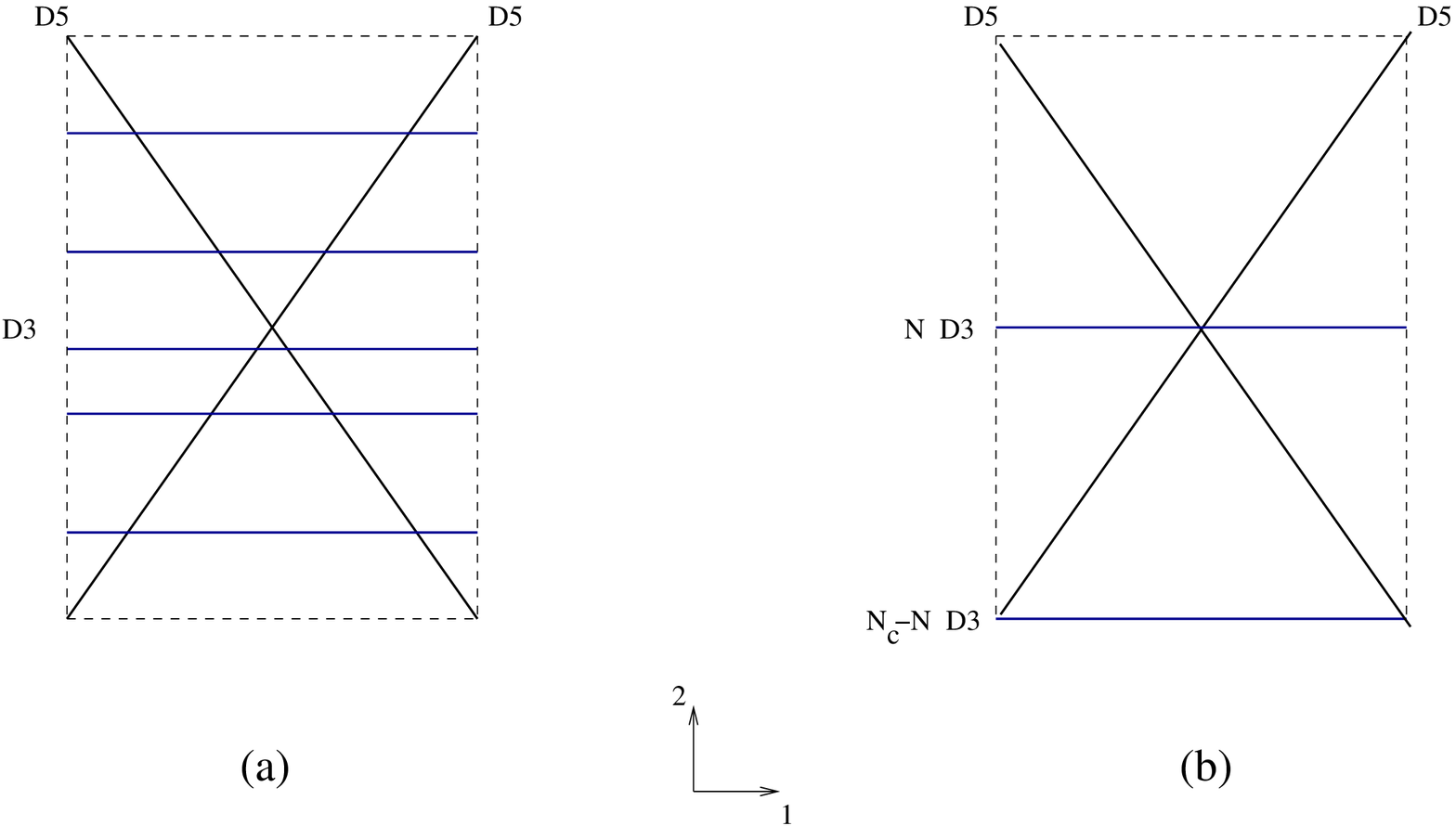}}

\bigskip

Classically, the potential on the Coulomb branch is flat, but quantum mechanically it is not. At a generic point in the Coulomb branch, the chiral superfields $Q^1$ and $\tilde Q_2$ are localized at different places on the $x^1$ circle (see figure 8 (a)). Locally in $x^1$ the theory thus looks like two copies of a $U(1)$ gauge theory, one with just $Q^1$ and $\Lambda^2$, the other with $\tilde\Lambda_1$, $\tilde Q_2$. Each of these theories separately breaks supersymmetry due to  a non-zero expectation value of the $D$ term in the quantum theory. Therefore, it is natural to expect that the theory that contains  both also has this property. A more detailed discussion of this issue is presented in appendix A.

The analysis of appendix A implies that the Coulomb branch of figure 8 (a) is replaced in the quantum theory by a discrete set of vacua, labeled by an integer $N$, which in the brane description corresponds to the number of color threebranes that are placed at one of the intersections of the flavor branes. The other $N_c-N$ color branes are placed at the other intersection (see figure 8 (b)). Superficially it looks like the integer $N$ runs from $0$ to $N_c$ but we shall next argue that it must satisfy further constraints.

The D-term conditions for the modes localized at one of the intersections in figure 8 (b), say the one with $N$ color $D3$-branes, are given in appendix A. They are the same as those for four dimensional $N=1$ SQCD with gauge group $U(N)$ and $N_f/2$ flavors $Q^1$, $\tilde Q_2$. The classical moduli space of that theory has been well studied (see \eg\ \refs{\IntriligatorAU,\TerningBQ}). For $N_f\ge 2N$, the gauge symmetry is generically completely broken, and the moduli space is $N_fN-N^2$ dimensional. For $N_f<2N$, the gauge symmetry is generically broken to $U(N-\half N_f)$ by the fundamentals. The classical Higgs moduli space is in this case $N_f^2/4$ dimensional and can be parametrized by the gauge invariant meson fields $Q^1\tilde Q_2$. In two dimensions there are also Coulomb moduli associated with the unbroken part of the gauge group, which break it further to the Cartan subalgebra. 

Quantum mechanically, we expect supersymmetry to be broken at intersections with $N_f<2N$. Indeed, in other situations of this sort, such as $N=1$ SQCD in four dimensions and $N=2$ SQCD in three dimensions with more colors than flavors, the theory develops a quantum superpotential for some of the classical moduli, that pushes them to infinity. We expect the same to happen here, but will not attempt to show that it does.

In the brane picture, the fact that the gauge symmetry is not completely broken along the Higgs moduli space implies that at a generic point in moduli space $N-\half N_f$ of the $D3$-branes continue to stretch between the $NS5$-branes. Quantum mechanically, such $D$-branes are typically repelled by other $D3$-branes ending on the fivebranes \GiveonSR. This generates a potential for the corresponding Coulomb branch moduli that pushes them away from the intersection. 

Assuming that the above picture is correct, we conclude that if we want the low energy theory to have a supersymmetric vacuum, $N$ must lie in the range
\eqn\rangennn{N_c-{N_f\over2}\le N\le {N_f\over2}~,}
where we also included the constraint that follows from requiring stability of the vacuum at the other intersection. Note that \rangennn\ implies in particular that $N_f\ge N_c$. This is satisfactory since if $N_f$ and $N_c$ are outside this range, the four dimensional theory does not have a vacuum even before we compactify it and turn on any background fields. 

To summarize, we conclude that quantum  vacua of the two dimensional theory obtained from $N=1$ SQCD via our construction are labeled by an integer $N$, which can be thought of as a discrete remnant of the classical Coulomb branch, and takes values in the range 
\eqn\rangenn{\max(0,N_c-\half N_f)\le N\le \min(N_c,\half N_f)~.}
The total number of disconnected branches of moduli space is
\eqn\numbcomp{N_{\rm br}=\cases{N_f-N_c+1&$\;\;{\rm for}\;\;N_f\le 2N_c$\cr
N_c+1&$\;\;{\rm for}\;\;N_f\ge 2N_c$\cr}}
In each component, the low energy theory consists of theories localized at the two intersections. In general, we expect these theories to be coupled by terms obtained from integrating out massive modes living on the threebranes and exchange of modes living on the fivebranes, but these couplings can be suppressed by taking the size to the $x^2$ circle to infinity. Therefore, below we shall assume that the two theories are decoupled. 

The bosonic part of the theory at the intersection corresponding to the $N$ color branes is a $\sigma$-model on the moduli space of solutions to the $U(N)$ D-term equations. As mentioned above, the complex dimension of this space is $N_fN-N^2$. The right-moving central charge, and the left-moving one that is equal to it due to the absence of a gravitational anomaly in the spectrum of Table 1, can be read off by going to large values of $Q$, $\tilde Q$, where the theory becomes weakly coupled. This gives 
\eqn\ccllrr{c_R=c_L=3(N_fN-N^2)~.}
Near the origin of the Higgs branch, the theory becomes strongly interacting in the IR due to the gauge dynamics. If $N_f\gg N$, one can use vector model techniques to solve it.

\bigskip

\ifig\loc{The magnetic brane configuration for general $N_f, N_c$.  Vacua are labeled by an integer $\hat N$ that runs over the range described in the text. For each $\hat N$ the theory splits into two decoupled theories at the intersections. }
{\epsfxsize2.1in\epsfbox{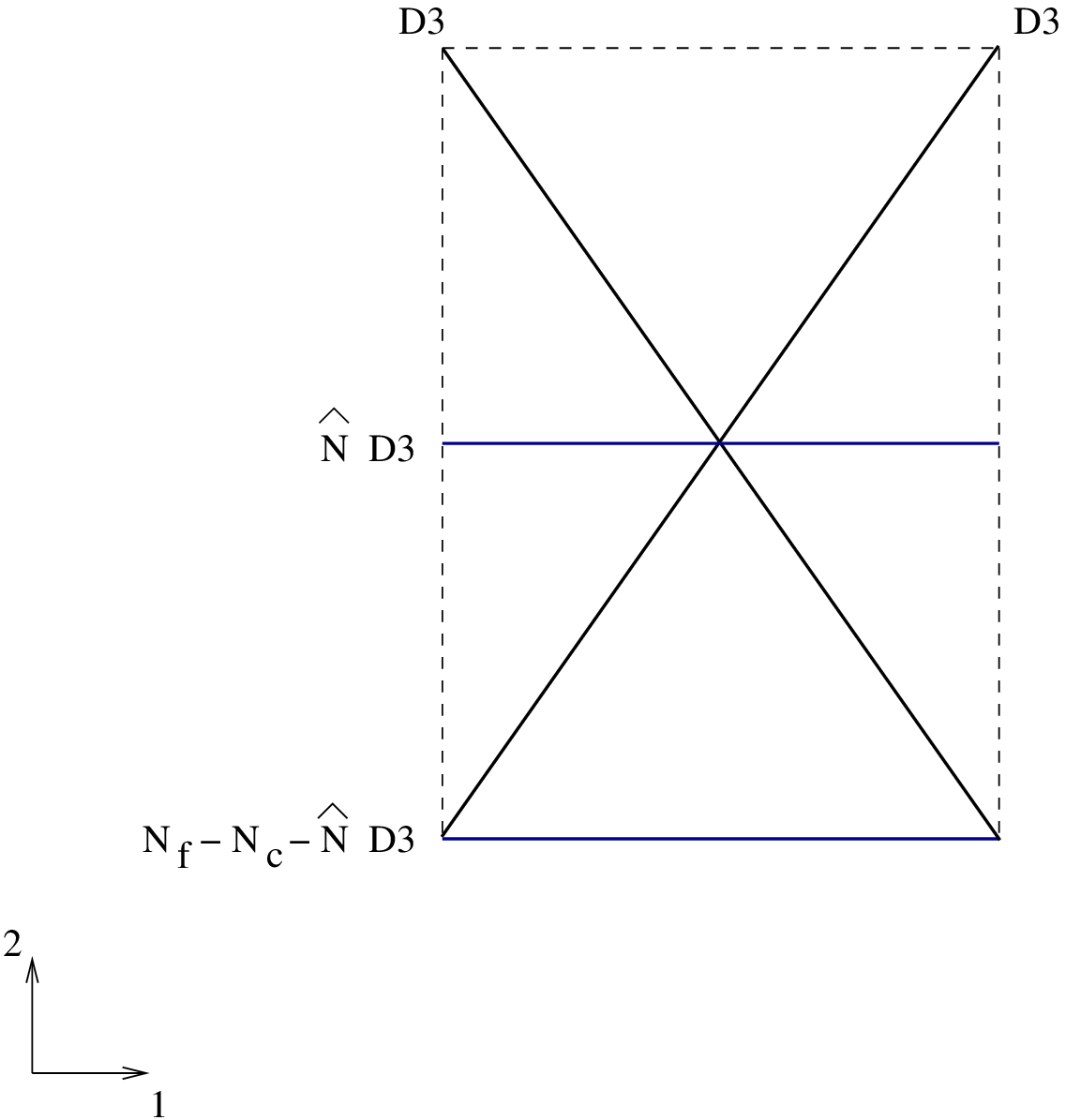}}

\bigskip

We next turn to the magnetic theory, whose brane description is given  in figure 9. Flavor $D3$-branes are plotted in black in the figure. They split into two groups according to the global $U(1)$ charges (which are $\pm 1$, as before), and intersect at two points on the torus. Color branes, plotted in blue, split into two groups of $\hat N$ and $N_f-N_c-\hat N$ which are placed at the two intersections, as in the electric theory.

The dynamics is again localized at the two intersections.  Consider \eg\ the intersection associated with the $\hat N$ color branes. It is a $U(\hat N)$ gauge theory with the matter content listed in Table 2 above and superpotential \zerotwospot. The parameter $\hat N$ takes value in the range $0\le \hat N\le N_f-N_c$, but as in the electric case we expect it to also satisfy the constraints $\hat N, N_f-N_c-\hat N\le N_f/2$. Indeed, if the singlet mesons $M^i_j$ were not coupled to the magnetic theory, the latter would be identical to the electric theory with the replacement $N_c\to N_f-N_c$, $N\to \hat N$, and the parameter $\hat N$ would satisfy the analog of \rangenn,
\eqn\rangenhat{\max(0,\half N_f-N_c)\le \hat N\le \min(N_f-N_c,\half N_f)~.}
The number of branches of moduli space would again be given by  \numbcomp, which is essentially the statement that this expression is invariant under Seiberg duality, $N_c\to N_f-N_c$. Coupling the meson fields to the magnetic theory is not expected to change the number of branches, hence the vacua of the full magnetic theory are also expected to be labeled by the integer $\hat N$ taking value in the range \rangenhat.

The map between the electric and magnetic vacua can be obtained by comparing the `t Hooft anomalies of the two models. In the electric theory, the spectrum of Table 1 gives rise to the following non-zero anomalies:
\eqn\nonzeroanom{\eqalign{
[SU(N_f/2)_1]^2:&\qquad +N\cr
[SU(N_f/2)_2]^2:&\qquad -N\cr
[SU(N_f/2)_3]^2:&\qquad -N\cr
[SU(N_f/2)_4]^2:&\qquad +N\cr
}}
In the magnetic theory we find the same result, with $N$ replaced by $N_f/2-\hat N$. Therefore, we conclude that the map between the electric and magnetic vacua is 
\eqn\mapnn{\hat N={N_f\over2}-N~.}
This relation maps the range \rangenn\ to \rangenhat. 

The equivalence between the electric and magnetic theories can be thought of as a strong-weak coupling duality in the following sense.
As explained above, the electric theory, which is a $(0,2)$ sigma model on the Higgs branch of the $U(N)$ gauge theory with $N_f/2$ flavors, becomes weakly coupled in the region of large $Q^1$, $\tilde Q_2$. In the magnetic theory, the field parametrizing the target space of this sigma model is the singlet meson $M^1_2$. Since it does not appear in the magnetic superpotential \zerotwospot, superficially it appears that this field is free in the infrared. However, one can show that integrating out the massive fields leads to the appearance of interactions of this field with the Fermi superfields that are charged under the gauge group in Table 2. Denoting these superfields collectively by $\lambda$, the leading couplings take the schematic form 
\eqn\coupmlambda{\CL\sim\int d^2\theta|M^1_2|^2|\lambda|^2~.}
These interactions make the magnetic theory strongly coupled in the large $M^1_2$ region, where the electric theory is weakly coupled. This is reminiscent of what happens in four dimensions \SeibergPQ, where going along the electric Higgs branch makes the electric (magnetic) theory more weakly (strongly) coupled.

Another sense in which the relation between the electric and magnetic theories is a strong-weak coupling duality is the following. As mentioned above, for $N_f/2\gg N$, the electric theory becomes a gauge theory with many more flavors than colors, which can be treated using large $N$ vector model techniques. In this sense, the electric gauge dynamics becomes weakly coupled in this limit. The relation \mapnn\ implies that in this limit the magnetic gauge dynamics is strongly coupled, since the rank of the magnetic gauge group, $\hat N$, is comparable to the number of flavors, $N_f/2$. Conversely, for $N_f/2\gg\hat N$, the magnetic gauge dynamics is weakly coupled, since the number of flavors is much larger than the rank of the magnetic gauge group, while the electric theory is strongly coupled since the electric rank and number of flavors are comparable. This is similar to what happens in the three dimensional analogs of Seiberg duality \refs{\AharonyGP,\GiveonZN}.

Earlier in this section we wrote down the central charge of the electric theory localized at an intersection with $N$ color branes, \ccllrr. This was done by studying the low energy theory at large values of the scalar fields that parametrize the Higgs branch, where the theory simplifies. In the magnetic theory, the analogous calculation is more involved, due to the presence of the singlet mesons and superpotential \zerotwospot. However, we can attempt to calculate the central charge by using its relation to the anomaly of the $U(1)_R$ symmetry which belongs to the $N=2$ superconformal multiplet. 

In the electric theory this $U(1)_R$ assigns charge zero to $Q$, $\tilde Q$, since they parametrize the target space of the low energy $\sigma$-model (as in \WittenYU). The fermions in these multiplets thus have R-charge $-1$. The Fermi superfields have R-charge zero, being left-moving. The adjoint chiral superfield $\Sigma$ is massive, hence its R-charge is one. Finally, $\Upsilon$ has R-charge one, as is clear from its mode expansion \upsilone. Altogether we find the anomaly
\eqn\elecanom{k=2\times\half N_fN-N^2~,}
with the first contribution coming from the fermions in $Q$, $\tilde Q$, and the second from $\Upsilon$. Multiplying by three we get the central charge of the theory, \ccllrr.   

In the magnetic theory, the adjoint fields $\Upsilon$ and $\Sigma$ have R-charge one, as before. The magnetic meson 
$M^1_2$ has R-charge zero, since it is related to the chiral superfield $Q^1\tilde Q_2$ in the electric theory. We will assume that the fields $M^1_1$ and $\Lambda^1_1$ are massive in the quantum theory. Indeed, a coupling of the E-type \dlambda, $\bar\DD_+\Lambda^1_1 = mM^1_1$ is consistent with all the symmetries of the problem, and we expect it to be generated by the quantum dynamics, since the electric field dual to $M^1_1$, $Q^1\tilde Q_1$, is massive. The singlet meson $\Lambda_1^1$ in the magnetic theory is dual to $Q^1\tilde\Lambda_1$ in the electric one, which has R-charge zero. Thus, we assign R-charge zero to $\Lambda^1_1$ and one to $M^1_1$. The magnetic chiral superfields $q_2, \tilde q^1$ are assigned charge $R_q$, while the Fermi superfields $\lambda_1$, $\tilde\lambda^2$ are assigned charge $R_\lambda$. 

The magnetic superpotential \zerotwospot\ then implies that $R_q+R_\lambda=0$, and $R(\Lambda^2_1)=1-2R_q$. Thus, all R-charges are determined by $R_q$. To find it we evaluate the $U(1)_R^2$ anomaly in the magnetic theory, which gives
\eqn\maganom{k=N_f\hat N(R_q-1)^2  - N_f \hat N R_q^2 + \frac{N_f^2}{4} - \frac{N_f^2}{4}(1-2R_q)^2 - \hat N^2}
and demand that it be equal to \elecanom. This gives\foot{A second solution, $R_q=1/2$, can be discarded based on the $U(1)_e U(1)_R$ anomaly.} 
\eqn\formqq{R_q=-R_\lambda={2N\over N_f}-\half={N-\hat N\over N_f}~.}
The electric theory has a non-vanishing $U(1)_eU(1)_R$ anomaly
\eqn\ueerr{U(1)_eU(1)_R:\;\; -NN_f~.}
One can check that the magnetic theory with the charge assignments above gives the same anomaly.

It is natural to ask whether the result \formqq\ can be understood directly in the magnetic theory, rather than by demanding agreement with the electric one. We do not have a full answer to this question, but can offer the following observations. First, note that the $U(1)_R$ anomaly \elecanom\ can be written in terms of magnetic variables as
\eqn\maganomaly{k=\left(N_f\over2\right)^2-\hat N^2}
If the $\hat N^2$ term was absent, it would be natural to interpret it as due to the $(N_f/2)^2$ chiral superfields $M^1_2$, which do not appear in the superpotential \zerotwospot, and thus, at least naively, seem to be free. In fact, as mentioned above, these fields have interactions with the Fermi superfields charged under the magnetic gauge group, of the schematic form \coupmlambda. These and other interactions presumably reduce the rank of the part of the $M^1_2$ matrix that is genuinely free by $\hat N$, and give rise to \maganomaly. We haven't studied the detailed way in which this happens, but it seems that at least in the limit $N_f\gg 2\hat N$, we can approximately treat the full ${N_f\over2}\times {N_f\over2}$ matrix $M^1_2$ as consisting of free non-compact chiral superfields, whose R-charge is fixed at zero. The R-charge $R_q$ can then be evaluated by extremizing $k$ \maganom\ subject to this constraint \refs{\BeniniCZ,\BeniniCDA}.\foot{For a discussion of $k$ extremization in the context of gauge/gravity duality, see \KarndumriIQA.} This gives $R_q=N/N_f$, which agrees with \formqq, since in the limit $N_f/2\gg\hat N$ one has $N=N_f/2-\hat N\simeq N_f/2$. 

For generic $N_f,\hat N$, it is less clear what to do. From the duality we expect part of the matrix $M^1_2$ to give rise to a non-compact moduli space of complex dimension \maganomaly, which implies that using $k$ extremization is more involved. The derivation of \formqq\ in that case remains an open problem.

\newsec{Discussion}

Our main purpose in this paper was to explore the theories one gets by starting with four dimensional $N=1$ supersymmetric gauge theories and reducing them on a two-torus with a non-zero magnetic field and corresponding D-term for a global $U(1)$ symmetry. This gives rise at long distances to two dimensional field theories with $(0,2)$ supersymmetry, and our goal was to explore their dynamics. We focused on the special case of $N=1$ SQCD and a particular choice of global $U(1)$ symmetry, which is a subgroup of the global symmetry group of the model.  We showed that the low energy theory has a non-trivial vacuum structure, and in a given vacuum it gives a $(0,2)$ supersymmetric $\sigma$-model on the Higgs branch. 

We also discussed the fate of Seiberg duality in four dimensions under such reduction. We presented evidence that it survives the compactification to two dimensions and reduction of supersymmetry from four to two supercharges. In particular, the vacuum structure of the electric theory appears to be reproduced by the magnetic theory. Our analysis was incomplete in many respects. In particular, parts of the discussion of quantum effects in section 7 relied to a large extent on the brane picture, and analogies to what happens in other dimensions. It would be interesting to fill these gaps and understand the dynamics directly in field theory. 

Most of our analysis was restricted to the case where the global charges of the fields under the $U(1)$ for which we turn the background fields are $+1$ for half of the flavors and $-1$ for the other half. One might wonder what happens if we take a more general choice of solution to the constraints \anomsqcd, \stronganom. 

A simple generalization is to replace the charges $\pm 1$ in the construction by $\pm n$, with $n$ an integer larger than one. It is particularly easy to see what happens in that case in the type IIB brane construction of section 6. The flavor branes have $2n$ intersections, which split into two groups of $n$ intersections at a given value of $x^2$. Thus, the low energy theory is a sum of $n$ theories of the sort described above with gauge group $U(N)$ living at one value of $x^2$, and $n$ theories with gauge group $U(N_c-N)$ living at the other. Thus, taking $n>1$ does not introduce any genuinely new elements into the discussion.    

Other generalizations of our construction involve non-trivial solutions of the equations for the $e_i$,  \anomsqcd, \stronganom. As a simple example, consider a $U(1)$ gauge theory with three flavors and charges $e_1=2, e_2=e_3=-1$ (and $\tilde e_i=- e_i$). It is easy to see what happens in this model using the brane construction. The three flavors correspond to $D5$-branes, one of which has winding number two (as in figure 5), and the other two winding one with the opposite orientation. The color $D3$-brane is again pushed towards the intersection of the $D5$-branes; however now even at the intersection, the D-term potential leads in the quantum theory to supersymmetry breaking due to an incomplete cancelation between the contributions of $Q$ and $\tilde Q$. This is typically what happens for general global charge assignments. 

There are many natural generalizations of the discussion in this paper. In particular, one can study more general $N=1$ gauge theories and the corresponding two dimensional $(0,2)$ models \KutasovHHA, generalize to models with $N=2$ SUSY in four dimensions, and embed our construction in a holographic setting (for $N=4$ SYM this was discussed in \refs{\AlmuhairiWS,\BeniniCDA}). Hopefully, such generalizations will help elucidate the relation between the dynamics of the underlying four dimensional models and the long distance two dimensional ones, and shed light on both.  
It would also be interesting to understand the relation between our construction and other ways of getting $(0,2)$ models in two dimensions from higher dimensional theories, such as \refs{\TongQJ,\ShifmanKJ,\GaddeLXA}.

\bigskip

\noindent{\bf Acknowledgements}: We thank  A. Giveon, Z. Komargodski, I. Melnikov, C. Quigley  and S. Sethi for discussions.   This work was supported in part by DOE grant DE-FG02-90ER40560, NSF Grant No. PHYS-1066293 and the hospitality of the Aspen Center for Physics, and by the BSF -- American-Israel Bi-National Science Foundation. The work of JL was supported in part by an NSF Graduate Research Fellowship.

\appendix{A}{Coulomb branch analysis}

In this appendix we discuss the classical Coulomb branch of the theory discussed in the text, and show that it is lifted by quantum corrections.

\subsec{Classical Analysis}
We begin by studying the classical vacuum structure. To introduce some of the issues in a simpler setting, we start with the case $N_f = N_c = 1$ (depicted in figure 5). This case does not satisfy the constraints \stronganom\ but is a useful warm-up for models with larger $N_f$ and $N_c$. 
Before turning on the background fields (in the brane realization when the threebrane and fivebrane in figure 5 are both stretched in the $x^1$ direction), the low energy theory is a three-dimensional field theory with $N=2$ SUSY, $U(1)$ gauge group, and two charged chiral superfields $Q, \tilde Q$ with global charges $+1$ and $-1$ respectively, compactified on a circle. After turning on the background fields, $Q$ and $\tilde Q$ give (0,2) chiral and Fermi superfields $Q, \tilde\Lambda$, while the vector superfield gives rise to the (0,2) field strength superfield $\Upsilon$ and adjoint chiral superfield $\Sigma$.

The classical moduli space consists of a one-dimensional Coulomb branch labeled by $\sigma$. This is easy to see in the brane picture, where the Coulomb branch is parametrized by the position of the $D3$-brane in the $x^2$ direction, and the dual of the $U(1)$ gauge field on the threebrane.   The two dimensional theory does not have a Higgs branch; the field $Q$ is set to zero by the D-term potential, which is proportional to $|Q|^4$. We would like to verify this picture in the higher dimensional field theory. 
 
The fields that enter our analysis are the scalar field $\phi = {\rm Im}\sigma$ parametrizing the position of the $D3$-brane in the $x^2$ direction, the auxiliary field $D$ in the vector multiplet on the $D3$-brane, and $Q$, as well as the background fields $\phi_B$ (which parametrizes the location of the $D5$-brane in $x^2$) and $D_B$. In addition, it is necessary to include in the discussion fields $Q_n$ that correspond to wound open strings in the brane picture.
The chiral superfield $Q$ is in the brane picture a string stretching from the threebrane to the fivebrane. Because $x^2$ is compact, there are actually infinitely many such fields $Q_n$, corresponding to strings that stretch between the threebrane and fivebrane while winding $n$ times around the $x^2$ circle.\foot{These fields owe their existence to the compactness of $x^2$. From the point of view of the underlying four dimensional theory they are the momentum modes of the four dimensional field $Q$ in the $x^2$ direction.} For a given position of the D-branes, at most one of these fields is light; however, relative motions of the branes change which one it is. For example, keeping the threebrane fixed and sending the fivebrane around the $x^2$ circle takes the mass of $Q_n$ to that of $Q_{n\pm1}$, depending on the direction of the motion.  In the presence of the non-trivial $\phi_B$ required by our construction, the position of the fivebrane in $x^2$ changes with $x^1$, so a similar shift occurs as a function of $x^1$.

Since we are only interested in the scalar potential, we take all fields to be constant in $(x^0,x^3)$. Denoting the remaining coordinate $x^1$ by $x$, the scalar potential is given by 
\eqn\threedtoy{
U = \int_0^{2\pi R_1}dx\left[\frac 1{2g^2}(\phi'^2-D^2)+\sum_{n\in Z}\left(|Q'_n|^2    - |Q_n|^2 (D_B-D) + |Q_n|^2(\phi_B-\phi+2\pi n R_2 T)^2 \right)\right]
}
where $T = 1/2\pi\alpha'$ is the string tension and the prime denotes differentiation in $x^1$. \threedtoy\ includes the kinetic terms for the various fields and the standard couplings \indlll\ of charged fields to the auxiliary fields in the vector multiplets. The relative minus sign between the $D_B$ and $D$ couplings is due to the fact that $Q$ corresponds to an oriented $3-5$ string, and therefore has opposite charges under the gauge groups on the $D3$ and $D5$-branes. 

The mass (last) term is the energy of a string stretching from the threebrane to the fivebrane while winding $n$ times around the $x^2$ circle. Note that we have set the $x$ component of the dynamical gauge field $A_1(x)$ to zero; this can (almost) be done by a $U(1)$ gauge transformation. The only information in $A_1$ that cannot be gauged away is the Wilson line $\exp(i\int dxA_1)$ or in other words constant $A_1$, which parametrizes one of the two directions of the compact Coulomb branch of the model (the other being the constant mode of $\phi$). We shall set it to zero below.

To arrive at the configuration in figure 5, we take the background fields to have the values
\eqn\bkgd{
\phi_B = B x, \qquad D_B = B~.
}
Plugging \bkgd\ into \threedtoy\ gives
\eqn\tdtoy{U = \int_0^{2\pi R_1} dx\left[ \frac 1{2g^2}(\phi'^2-D^2)+\sum_n\left(|Q_n' + (\phi_B  - \phi+2\pi n R_2 T )Q_n|^2  -
|Q_n|^2(\phi'-D)\right) 
\right]~,
}
where we used integration by parts and the fact that $\phi_B'=D_B$. 

We also require that the integrand \tdtoy\ be well defined on the circle. 
While $\phi$ and $D$ are periodic, $\phi_B$ satisfies $\phi_B(x+2\pi R_1)=\phi_B(x)+2\pi R_1 B$. If $B$ satisfies 
\eqn\quantbb{R_1B=lR_2T}
for some integer $l$, we can absorb this violation of periodicity by demanding that 
\eqn\perqn{Q_n(x+2\pi R_1)=Q_{n+l}(x)~.}
The constraint \quantbb\ on the magnetic field $B$ is the same as the one obtained in the four dimensional analysis \dirquant. The two are related by the standard T-duality relation $R_2^{(IIB)}=\alpha'/R_2^{(IIA)}$. In the following we shall restrict to the case $l = 1$. 

The equation of motion for $D$ in \tdtoy\ sets it to 
\eqn\formdddd{D=g^2\sum_n|Q_n|^2~.}
Plugging this back into \tdtoy\ yields the energy function 
\eqn\tdtoynew{U = \int_0^{2\pi R_1} dx\left[ \frac 1{2g^2}\left(\phi'-g^2\sum_n|Q_n|^2\right)^2+\sum_n|Q_n' + (\phi_B  - \phi+2\pi n R_2 T )Q_n|^2
\right].
}
Note that the first term in \tdtoynew\ is the square of the supersymmetry variation of the right-moving gaugino $\lambda_+$ in the vector multiplet, and the second term is the square of the variation of the right-moving fermion $(\psi_+)_n$ in the chiral multiplet $Q_n$. 
 
The vanishing of the energy necessary for supersymmetry implies that the two terms in \tdtoynew\ vanish separately. In particular,
\eqn\formphi{\phi'=g^2\sum_n|Q_n|^2.} 
However, this relation is inconsistent with the periodicity of $\phi$. Indeed, upon integrating \formphi\ over the $x^1$ circle, the l.h.s. vanishes since $\phi$ is periodic, while the r.h.s. is positive definite. We conclude that $Q_n$ must vanish for the vacuum to be supersymmetric. 
On the other hand, configurations with $Q_n = \phi' = 0$ manifestly satisfy $U=0$ for arbitrary (constant) values of $\phi$.
The classical moduli space of the model with $G=U(1)$ and one flavor therefore indeed consists of a Coulomb branch parametrized by $\phi$.

To get a non-trivial Higgs branch, we add another chiral superfield with opposite $U(1)$ charge, so that there are two flavors $(Q^i, \tilde Q_i)$, $i = 1,2$. This is the simplest model that satisfies \stronganom. After turning on the background fields \bkgd, we arrive at the system of figure 6. The brane picture suggests that there is again a Coulomb branch, parametrized by the scalar field $\sigma$, whose imaginary part is the position of the $D3$-brane in $x^2$. At generic points on the Coulomb branch, the chiral superfields $(Q^i,\tilde Q_i)$ give rise to two $(0,2)$ chiral superfields $Q=Q^1$ and $\tilde Q=\tilde Q_2$ (as well as two Fermi superfields, $\Lambda^2$ and $\tilde\Lambda_1$), which are localized at different points on the $x^1$ circle.
 
To analyze this case, we again start from the three-dimensional Lagrangian and turn on the background fields. The analogs of \formdddd, \tdtoynew\  are
\eqn\tdtoytwo{\eqalign{D=&g^2\sum_n\left(|Q_n|^2-|\tilde Q_n|^2\right)~,\cr
U =& \int_0^{2\pi R_1} dx\left\{ \frac 1{2g^2}\left[\phi'-g^2\sum_n\left(|Q_n|^2-|\tilde Q_n|^2\right)\right]^2\right.\cr
+&\left.\sum_n\left[|Q_n' + (\phi_B  - \phi+2\pi n R_2 T )Q_n|^2+|\tilde Q_n' + (\phi_B  + \phi+2\pi n R_2 T )\tilde Q_n|^2\right]
\right\}~.
}}
The three terms in the scalar potential correspond to the supersymmetry variations of the right-moving fermion inside the vector, $Q_n$ and $\tilde Q_n$ multiplets respectively, and all three must vanish for (0,2) supersymmetry to be preserved. It is clear that $U=0$ when $\phi' = Q_n = \tilde Q_n = 0$ for arbitrary values of constant $\phi$, corresponding to the classical Coulomb branch. One can also satisfy the conditions for unbroken supersymmetry by taking $Q_n(x) = \tilde Q_n(x)$ and $D = \phi' = 0$; this is the Higgs branch mentioned above. 

Note that for constant $\phi$ the condition for vanishing of the last two terms in $U$ can be thought of as follows. Using the matching conditions \perqn\ (with $l=1$), we can construct out of the $Q_n$ a single function $Q(x)$ on the covering space of the $x^1$ circle, and similarly for $\tilde Q$. Vanishing of the last two terms in $U$ is the requirement that $Q(x), \tilde Q(x)$ are ground states of the harmonic oscillator, localized at $x = \phi/B$ and $-\phi/B$ respectively. The first term in $U$ \tdtoytwo\ then requires $\phi=0$, \ie\ it fixes the Coulomb modulus.

In the classical theory there actually seem to be solutions with $U=0$ that have non-trivial $D = \phi' $ (as well as $Q, \tilde Q \neq 0)$. We shall not discuss them here since as we shall see below, they are lifted by quantum effects.

The above picture can be generalized to higher $N_f, N_c$. 
The analog of \tdtoytwo\ for the general case is
\eqn\threedtwo{\eqalign{D^a=&g^2\sum_n\left(\bar Q_n^i T^aQ_n^i - \bar{\tilde Q^n_\rho} T^a\tilde Q^n_\rho\right) - g^2 [\bar\sigma, \sigma]^a ~,\cr
U =& \int_0^{2\pi R_1} dx\left\{ \frac 1{2g^2}\left[(\phi^a)'-g^2\sum_n\left(\bar Q_n^i T^aQ_n^i - \bar{\tilde Q^n_\rho} T^a\tilde Q^n_\rho\right)\right]^2\right.\cr
+&\left.\sum_n\left[|(Q^i_n)' + (\phi_B  - \phi_aT^a+2\pi n R_2 T )Q^i_n|^2+|(\tilde Q^n_\rho)' + (\phi_B  + \phi_aT^a+2\pi n R_2 T )\tilde Q_\rho^n|^2\right]
\right\}~.
}}
A new element in this case is the fact that the scalar fields $\phi^a$ belong to the adjoint representation of  $U(N_c)$. We can think of them as forming a Hermitian $N_c \times N_c$ matrix, which can be diagonalized by a $U(N_c)$ gauge transformation. The eigenvalues of $\phi$ parametrize the Coulomb branch of the model and correspond in the brane desription to the positions of the $N_c$ $D3$-branes in the $x^2$ direction (as in figure 8 (a)). It is easy to generalize the $U(1)$ analysis to this case. 

\subsec{Quantum analysis}

We now turn to a discussion of quantum effects. Since we are interested in the Coulomb moduli we set $Q_n = \tilde Q_n = 0$ and ask whether the moduli space labeled by $\sigma$ survives in the quantum theory. To see the basic physics it is enough to consider the case $N_c=1$. The generalization to larger $N_c$ is straightforward.

The scalar potential $U$ for this case can be written in the form %
\eqn\tdtoytt{\eqalign{
U &= \int_0^{2\pi R_1} dx \left[\frac{1}{2g^2}(\phi'^2-D^2)+ \sum_n(|Q'_n + (\phi_B - \phi + 2\pi n R_2T)Q_n|^2 \right. \cr &
\left. + |\tilde Q_n' + (\phi_B + \phi + 2\pi n R_2T)\tilde Q_n|^2 - (|Q_n|^2 - |\tilde Q_n|^2)(\phi'-D)) \right],
}}
where we suppressed the flavor indices of $Q$, $\tilde Q$. Although we are interested in the fate of classical vacua with $Q=\tilde Q=0$, in the quantum theory we need to include the effects of fluctuations of these fields on the Coulomb modulus $\phi$. At large $N_f$ this can be done by performing the Gaussian integral over $Q$, $\tilde Q$ and treating $\phi$ and $D$ semi-classically, and we shall do that below. In general $\phi$ and $D$ should be path integrated over as well, but we do not expect this to change the conclusions. 

The fields $Q, \tilde Q$ have an expansion in Landau levels,
\eqn\defqn{
Q_n(x^0, x^3, x) = \sum_{a=0}^\infty Q_a(x^0, x^3)\varphi_n^a(x), \qquad \tilde Q_n(x^0, x^3, x) = \sum_a\tilde Q_a(x^0, x^3)\tilde\varphi^a_n(x)
}
where the $Q_a, \tilde Q_a$ are two-dimensional fields with masses that scale as $M^2_a = a|eB|$ \shmspectrum.
At one loop one has (see \eg\ \TongQJ\ for a related discussion)
\eqn\ovev{
\langle\bar Q_0Q_0\rangle = \langle\bar{\tilde Q_0}\tilde Q_0\rangle  ={N_f\over2} \int\frac{d^2k}{(2\pi)^2}\frac{1}{k^2+\mu^2} \sim {N_f\over2}\ln \frac{\Lambda}{\mu}
}
where $\mu$ is an infrared scale; for the massive fields $Q_{a\neq 0}, \tilde Q_{a\neq 0}$, $\mu$ is replaced by $m_a$. Since the infrared scale $\mu$ is in general well below $m_a$, the contributions of the massive fields to the two point function are suppressed relative to the massless ones. Therefore, we need only consider the fluctuations of $Q_0, \tilde Q_0$ in \tdtoytt. As discussed above, the wavefunctions $\varphi^0_n, \tilde\varphi^0_n$ are segments of the wavefunction of the ground state of the harmonic oscillator, localized at $x = \phi_0/B$ and $-\phi_0/B$ respectively.

It is conveninent to define $\phi_{nz} = \phi - \phi_0$, where $\phi_0$ is the average value of $\phi$ on the $x^1$ circle, and $\phi' = \phi'_{nz}$.  
After integrating out $Q_0, \tilde Q_0$ \ovev, the effective potential for $\phi_{nz}$ and  $D$ reads
\eqn\tdeff{\eqalign{
U_{\rm eff} &= \int_0^{2\pi R_1} dx \left[\frac 1{2g^2}(\phi'^2_{nz} - D^2) + {N_f\over2}\left(\ln\frac\Lambda\mu \right)|\phi_{nz}|^2\left(F_Q(x) + F_{\tilde Q}(x)\right)\right.
\cr &- \left. {N_f\over2}\left(\ln\frac\Lambda\mu\right) (F_Q(x) - F_{\tilde Q}(x))(\phi'_{nz} - D)\right].
}}
Here $F_Q(x)  = \sum_n |\varphi^0_n(x)|^2$ and $F_{\tilde Q}(x) = \sum_n|\tilde\varphi^0_n(x)|^2$. The two differ by a shift $F_{\tilde Q}(x) = F_Q(x + x_0)$ with $x_0$ proportional to the Coulomb modulus $\phi_0$. 

The equations of motion for $\phi'_{nz}$ and $D$ that follow from \tdeff\ are
\eqn\effeom{\eqalign{
\frac{1}{g^2}\phi''_{nz} -{N_f\over2} \left(\ln\frac\Lambda\mu \right)(F'_Q- F'_{\tilde Q}) &= N_f\left(\ln\frac\Lambda\mu \right)\phi_{nz}(F_Q + F_{\tilde Q}) \cr
\frac{1}{g^2}D - {N_f\over2}\left(\ln\frac\Lambda\mu\right)(F_Q - F_{\tilde Q}) &= 0.
}}
As mentioned previously, $\phi' - D$ is the variation of the right-moving gaugino in the vector multiplet which must vanish in order for (0,2) supersymmetry to be preserved. From the above equations of motion, one can show that this is equivalent to requiring $F_Q = F_{\tilde Q}$. But $F_Q, F_{\tilde Q}$ are identical up to a shift in $x$. This condition is satisfied iff the shift is zero, \ie, $Q, \tilde Q$ are localized at the same position in $x$.

To summarize, we find that for $N_c = 1$ the Coulomb moduli space labeled by $\phi$ is replaced in the quantum theory by two isolated vacua. In the brane picture, they correspond to having a straight $D3$-brane pass through one of the two intersections of the $D5$-branes.

In the non-Abelian case, the same mechanism leads to the collapse of the classical moduli space of figure 8 (a) to the isolated vacua of figure 8 (b). The dynamics at a given intersection involves additional phenomena that place further constraints on the integer $N$, as described in the text.

The above analysis involved in an important way the massive KK modes of the various three dimensional fields. It is natural to ask how the quantum effects that fix the Coulomb moduli manifest themselves in the two dimensional low energy theory of the light modes. We have not understood this in detail, but D-term couplings of the adjoint chiral superfield $\Sigma$ with the charged Fermi superfields, which can be shown to have the schematic form $\int d^2\theta|\Sigma|^2|\Lambda|^2$ (for small $\Sigma$) seem to play an important role in this problem.

\listrefs
 
\bye